\documentclass[showpacs, amsmath,amssymb, aps,showkeys]{revtex4}
\usepackage{amsmath,amssymb,amsfonts,latexsym,float,graphics,epsfig,bm}
\usepackage{graphicx}

\catcode`@=11 \@addtoreset{equation}{section}

\newcommand{\be}{\begin{equation}}
\newcommand{\ee}{\end{equation}}

\begin{document}
\title{Axisymmetric Solutions to Einstein Field Equations via Integral Transforms}
\author{D. Batic}
\email{davide.batic@ku.ac.ae}
\affiliation{
Department of Mathematics,\\  Khalifa University of Science and Technology,\\ Main Campus, Abu Dhabi,\\ 
United Arab Emirates}

\author{N. B. Debru}
\email{100049891@ku.ac.ae}
\affiliation{
Department of Physics,\\  Khalifa University of Science and Technology,\\ Main Campus, Abu Dhabi,\\ 
United Arab Emirates}

\author{M. Nowakowski}
\email{mnowakos@uniandes.edu.co}
\affiliation{
Departamento de Fisica,\\ Universidad de los Andes, Cra.1E
No.18A-10, Bogota, Colombia
}
\email{marek.nowakowski@ictp-saifr.org}
\affiliation{
ICTP South American Institute for Fundamental Research,\\ Rua Dr. Bento Teobaldo Ferraz 271, 
01140-070 S\~o Paulo, SP Brazil 
}

\date{\today}
\begin{abstract}
In this paper, we present new axisymmetric and reflection symmetric vacuum solutions to the Einstein field equations. They are obtained using the Hankel integral transform method and all three solutions exhibit naked singularities. Our results further reinforce the importance and special character of axisymmetric solutions in general relativity and highlight the role of integral transforms methods in solving complex problems in this field. We compare our results to already existing solutions which exhibit the same type of singularities. In this context we notice that most known axial-symmetric solutions possess naked singularities. A discussion of characteristic features of the newly found metrics, e.g., blueshift and the geometry of the singularities, is given.

\end{abstract}


\maketitle
\section{Introduction}
The Birkhoff theorem, as presented in \cite{Birkhoff,Jebsen,Israel}, guarantees the uniqueness of the spherical symmetric solutions of the Einstein field equations. A corresponding theorem (or a classification scheme) for axisymmetric solutions does not exist. As a result, we find numerous nonequivalent solutions \cite{book}. We draw attention to the standout among axisymmetric solutions: the Kerr metric \cite{Kerr}. It represents an axially symmetric rotating black hole with mass $M$ and angular momentum $J$, and is typically expressed in Boyer-Lindquist coordinates. It possesses a horizon that conceals all associated singularities.  Another notable exact solution is the Tomimatsu-Sato (TS) metric, which describes the geometry around a deformed
spinning mass with a deformation parameter $\delta =2$ \cite{tomsato,tomsatop}. Investigations into this metric have uncovered directional naked singularities, which are deemed unphysical \cite{gary,DB}. Unlike the Kerr metric, the TS-metric does not have a horizon to shield these curvature singularities. A third example worth mentioning is the so-called Majumdar-Papapetrou (MP) metric \cite{MP,Papas}, which is discussed in detail in \cite{HH}. The authors of that study conclude that {\bf{aside}} from the scenario of several black holes aligned in equilibrium, all other solutions using the MP ansatz exhibit singularities. This highlights the challenge of obtaining a physically plausible axially symmetric solution devoid of naked singularities, with the Kerr metric (and potentially some undiscovered examples) being exceptions. Given the absence of overarching theorems, identifying new axisymmetric solutions is crucial. Should the majority of them harbor naked singularities, such a state of affair would {elevate the few examples that have an event horizon. The void left by the absence of the Birkhoff theorem might then be filled by these multiple counter-examples.  

Conversely, naked singularities no longer appear to be the ``enfants terrible'' of General Relativity. It is widely acknowledged that Stephen Hawking lost a bet regarding naked singularities, having wagered against their existence. This bet was based on a proposal by Roger Penrose, who introduced the so-called ``Cosmic Censorship Hypothesis'' \cite{Roger}. This hypothesis posits that naked singularities cannot form and that all curvature singularities must be concealed by an event horizon. 

As appealing as such a conjecture might appear, deviations were predicted as early as the mid-1970s, when \cite{Sz} discovered that the quasi-spherical gravitational collapse of dust clouds could lead to the formation of naked singularities. In the 1980s, a series of pivotal papers \cite{dem1,dem2} identified a breach of the Cosmic Censorship in the gravitational collapse of a dust cloud but also explored the gravitational collapse of a self-gravitating scalar field, establishing that, under specific conditions, a naked singularity might emerge. Further, in the 1990s, studies \cite{Shapiro, Shapiro2} provided numerical evidence suggesting that singularities could arise during the gravitational collapse of  collisionless gas spheroids. Specifically, when the spheroids are compact enough, the curvature singularities reside behind black hole horizons. Yet, for sufficiently large spheroids, these singularities remain exposed, unhindered by event horizons. Given such findings, it is unsurprising that Hawking conceded his earlier bet, spurring deeper investigation into the nature of singularities and the boundaries of General Relativity. Subsequent research delved into potential infractions of the Cosmic Censorship Hypothesis \cite{Singh,Pen} and circumstances leading to the manifestation of naked singularities. For instance, work from \cite{Ori1,Ori2} revealed that naked singularities could emerge from self-similar  spherical gravitational collapse, with their structure further analysed in \cite{Joshi1,Joshi2}. Insights into the appearance of such singularities in spherical symmetric gravitational collapse with tangential pressure, or in the context of a perfect fluid, were discussed in \cite{Dw,Singh1,Harada1,Magli}. Moreover, \cite{Harada2} identified naked singularity formation in the collapse of a spherical cloud of counter rotating particles. Building on earlier results \cite{Ch1,Ch2}, findings by \cite{Hamade} highlighted the formation of naked singularities in the spherically symmetric collapse of a self-gravitating massless scalar field. An enlightening study \cite{Joshi3} determined that when strong shearing effects occurred near the singularity, an apparent horizon formation could be delayed, revealing the curvature singularity to external observers. Notably, naked singularities have been identified in Szekeres space-times, which are solutions to the Einstein field equations (EFEs) generated by irrotational dust \cite{Joshi4}. Additionally, \cite{Sandip} introduced an intriguing proposal: naked singularities might be potential candidates for Gamma-ray bursters. Studies on the emergence of naked curvature singularities in the Einstein-Gauss-Bonnet gravity and the Brans-Dicke Theory are covered in \cite{Maeda,Amir,Rudra}. For contemporary perspectives on the (in)stability of naked singularities we direct readers to \cite{Gibbons1,Gleiser,Cardoso,Dotti1,Dotti2,Dotti3,Dotti4}. 

Comparing these studies is a daunting endeavor, as they delve into various facets of naked singularities and breaches of the Cosmic Censorship Hypothesis. Yet, a clear distinction emerges. Some studies specifically address the genesis of naked singularities through the gravitational collapse of different matter forms, such as scalar fields, dust clouds or massive stars. Others concentrate on the broader ramifications of these singularities for our comprehension of fundamental physics, touching on aspects like the stability of event horizons or radiation generation. Given the extended body of literature on the subject, the significance of naked singularities within General Relativity is undeniable. Their existence would directly challenge the Cosmic Censorship Hypothesis. Furthermore, the observable effects of these singularities on surrounding matter and radiation could call into question the foundational tenets of General Relativity itself. Hence, probing the nature of naked singularities is paramount to delineate the boundaries, and potential shortcomings, of general relativity in depicting our physical universe. It is in this regard that that our current work gains its relevance. We begin with a broad-based ansatz for an axisymmetric metric in Weyl coordinates, transforming the Ernst equation into a Laplace equation. Employing the Hankel integral transform directly, we derive three novel solutions to the EFEs. All of them exhibit naked singularities. Moreover, two of the metrics we obtained,  are notable for approximating the Minkowski metric at space-like infinity. The discovery of new solutions to the EFEs featuring naked singularities is crucial, as it deepens our understanding of gravity and space-time under extreme circumstances. Beyond that, they can serve as pivotal tools for testing and refining quantum gravity theories, which aspire to bridge the gap between general relativity and quantum mechanics. Not to forget, naked singularities are theorized to influence the formation of black holes, which are one of the most exotic and fascinating objects in the universe. Therefore, comprehending the traits of naked singularities and their genesis offers profound insights into the broader cosmic picture.

The paper is organised as follows. In Section 2, we establish our
notations and conventions. Additionally, using a metric ansatz in the Weyl-Lewis-Papapetrou form, we briefly detail the simplification of the EFEs down to the Ernst equation. This is subsequently transformed into a homogeneous Laplace equation via the Weyl approach. In Section 3, the Hankel transform is extensively employed to produce new axisymmetric metrics that asymptotically approach the Minkowski at infinity. Where applicable, by inspecting the Newtonian gravitational potential linked to the metric coefficient $g_{00}$, we also aim to provide a physical interpretation of our results. In Section 4, we draw our conclusions and discuss future research directions related to naked singularities.

\section{Axisymmetric solutions from the Ernst potential}
The general form of a metric corresponding to axisymmetric solutions can be expressed in cylindrical coordinates $(x^0,x^1,x^2,x^3)=(t,\rho,z,\phi)$ as the Lewis-Papapetrou line element \cite{Lewis,Papa}
\begin{equation}\label{Lewis-Papa}
ds^2=f dt^2-2\kappa dtd\varphi-\ell d\varphi^2-e^\mu(d\rho^2+dz^2),
\end{equation}
where the unknown functions $f,\kappa,l,\mu$ are dependent on $\rho$ and $z$.  To ensure that the metric above reduces to the Minkowski metric at large distances, we impose that (\ref{Lewis-Papa}) takes the form
\begin{equation}\label{Minkowski}
ds^2=dt^2-\rho^2 d\varphi^2-d\rho^2-dz^2,
\end{equation}
as $\rho,z\to\infty$. From this, we deduce that $f\to 1$, $\kappa,\mu\to 0$ and $\ell\to\rho^2$ for a valid metric of the form (\ref{Lewis-Papa}). A prominent example of this general form is the Kerr metric, which describes the geometry around an uncharged axially symmetric rotating black hole characterized by mass $M$ and angular momentum $J$. It is typically represented in Boyer-Lindquist coordinates \cite{Kerr}. Another exact solution of notable interest is the Tomimatsu-Sato (TS) metric, which describes the geometry around a deformed spinning mass with a deformation parameter $\delta=2$ \cite{tomsato,tomsatop}. Studies of this geometry have unveiled ring-like naked singularities, which are predominantly regarded as unphysical \cite{gary}. Nevertheless, it is worth highlighting that solutions of this kind have been exclusively examined in the prolate spheroidal coordinate system. This raises an intriguing question: do analogous solutions in alternate coordinate systems retain comparable geometrical properties? This particular aspect will be the subject of future investigations. To derive new axisymmetric solutions, it is convenient to recast equation (\ref{Lewis-Papa}) into the Weyl-Lewis-Papapetrou form. This can be achieved by making the substitution $w=\kappa/f$, and recognizing that the unknown functions $f$, $\kappa$ and $\ell$  are interrelated through the equation $\kappa^2+f\ell=\rho^2$. The resulting form is then
\begin{equation}\label{Weyl-Lewis-Papa}
ds^2=f(dt-wd\varphi)^2-\frac{\rho^2}{f}d\varphi^2-e^\mu(d\rho^2+dz^2).
\end{equation}
When one attempts to solve the vacuum EFEs given by
\begin{equation}\label{Ricci-EFE}
R_{\alpha\beta}=0
\end{equation}
{with respect to the unknown functions appearing in the line element (\ref{Weyl-Lewis-Papa}), it emerges that the only non-vanishing components of the Ricci tensor are $R_{00}$, $R_{03}$, $R_{11}$, $R_{12}$, $R_{22}$, and $R_{33}$. By expanding the EFEs for these specific components, we can further reduce (\ref{Ricci-EFE}) into a system of coupled PDEs as follows
\begin{eqnarray}
&&f\left(\partial_{\rho\rho}f+\partial_{zz}f+\frac{\partial_\rho f}{\rho}\right)-\left(\partial_\rho f\right)^2-\left(\partial_z f\right)^2+\frac{f^4}{\rho^2}\left[\left(\partial_\rho w\right)^2+\left(\partial_z w\right)^2\right]=0,\label{2.12a}\\
&&f\left(\partial_{\rho\rho}w+\partial_{zz}w-\frac{\partial_\rho w}{\rho}\right)+2\left(\partial_\rho w\partial_\rho f+\partial_z w\partial_z f\right)=0,\label{2.12b}\\
&&\partial_\rho\mu=-\frac{\partial_\rho f}{f}+\frac{\rho}{2f^2}\left[\left(\partial_\rho f\right)^2-\left(\partial_z f\right)^2\right]-\frac{f^2}{2\rho}\left[\left(\partial_\rho w\right)^2-\left(\partial_z w\right)^2\right],\label{2.13a}\\
&&\partial_z\mu=-\frac{\partial_z f}{f}+\frac{\rho}{f^2}\partial_\rho f\partial_z f-\frac{f^2}{\rho}\partial_\rho w\partial_z w.\label{2.13b}
\end{eqnarray}
Interestingly, from the above, we observe that the equation $\partial_\rho(\mathcal{A}\partial_\rho w)+\partial_z(\mathcal{A}\partial_z w)=0$, where $\mathcal{A}=f^2/\rho$, aligns with (\ref{2.12b}). This suggests the construction of a function $u=u(\rho,z)$, fulfilling the conditions
\begin{equation}\label{3.2}
\partial_\rho u=\frac{f^2}{\rho}\partial_z w,\quad
\partial_z u=-\frac{f^2}{\rho}\partial_\rho w.
\end{equation}
Such an approach allows the rewriting of equations (\ref{2.12a}) to (\ref{2.13b}) in the following form, namely
\begin{eqnarray}
f\nabla^2 f&=&\left(\partial_\rho f\right)^2+\left(\partial_z f\right)^2-\left[\left(\partial_\rho u\right)^2+\left(\partial_z u\right)^2\right],\label{3.3a}\\
f\nabla^2 u&=&2\left(\partial_\rho f\partial_\rho u+\partial_z f\partial_z u\right),\label{3.3b}\\
\partial_\rho\left(\mu+\ln{f}\right)&=&\frac{\rho}{2f^2}\left[\left(\partial_\rho f\right)^2-\left(\partial_z f\right)^2\right]+\frac{\rho}{2f^2}\left[\left(\partial_\rho u\right)^2-\left(\partial_z u\right)^2\right],\label{3.4a}\\
\partial_z\left(\mu+\ln{f}\right)&=&\frac{\rho}{f^2}\left(\partial_\rho f\partial_z f+\partial_\rho u\partial_z u\right).\label{3.4b}
\end{eqnarray}
Here, the Laplace operator in cylindrical coordinates is represented as $\nabla^2=\rho^{-1}\partial_\rho(\rho\partial_\rho\cdot))+\rho^{-2}\partial_{\varphi\varphi}+\partial_{zz}$. Interestingly, one can recognize (\ref{3.3a}) and (\ref{3.3b}) {\bf{as}} the real and imaginary parts of the Ernst equation \cite{Ernst}. In cylindrical coordinates, the Ernst equation is a complex second order, nonlinear  PDE given by 
\begin{equation}\label{Ernst}
\Re{\left(\mathcal{E}\right)}\nabla^2\mathcal{E}=\left(\partial_\rho\mathcal{E}\right)^2+\left(\partial_z\mathcal{E}\right)^2,\quad\Re{\mathcal{E}}=f, \quad\Im{\mathcal{E}}=u,
\end{equation}
where $\mathcal{E}=f+iu$ and $\Re,\Im$ denote as usual the real and imaginary parts of the complex-valued function $\mathcal{E}$. By making use of the ansatz
\begin{equation}\label{ansatz-phi}
\mathcal{E}=\frac{\Phi-1}{\Phi+1},
\end{equation}
with $\Phi$ being a yet undetermined complex-valued function, we can reformulate the Ernst equation  as
\begin{equation}\label{Ernst-cylind}
\left(|\Phi|^2-1\right)\nabla^2\Phi=2\Phi^{*}\left[\left(\partial_\rho\Phi\right)^2+\left(\partial_z\Phi\right)^2\right].
\end{equation}
Historically, the TS metric is derived by solving this form of the Ernst equation in prolate spheroidal coordinates \cite{tomsato,tomsatop}. For our investigation, the focus remains on the investigation of axially symmetric exact solutions to (\ref{Ricci-EFE}) in the so-called Weyl coordinates $(\rho,z)$. In this coordinate system, the Laplace operator simplifies to $\nabla^2=\rho^{-1}\partial_\rho(\rho\partial_\rho\cdot))+\partial_{zz}$. On introducing an ansatz of the form $\Phi(\rho,z)=e^{-i\alpha}F(\Psi(\rho,z))$ with $\alpha\in\mathbb{R}$, it is possible to choose $F$ such that the Ernst equation reduces to the Laplace equation. More precisely, we find that
\begin{equation}
(F^2-1)\frac{dF}{d\Psi}\nabla^2\Psi+\left[(F^2-1)\frac{d^2F}{d\Psi^2}-2F\left(\frac{dF}{d\Psi}\right)^2\right]\left[\left(\frac{d\Psi}{d\rho}\right)^2+\left(\frac{d\Psi}{dz}\right)^2\right]=0.
\end{equation}
It is evident that $\Psi$ satisfies the Laplace equation
\begin{equation}
\nabla^2\Psi=0
\end{equation}
under the condition
\begin{equation}
(F^2-1)\frac{d^2F}{d\Psi^2}-2F\left(\frac{dF}{d\Psi}\right)^2=0.
\end{equation}
The general solution to this equation is given by
\begin{equation}\label{gen_f}
F(\Psi)=\pm\frac{c_2 e^{2c_1\Psi}+1}{c_2 e^{2c_1\Psi}-1},
\end{equation}
where $c_1$ and $c_2$ represent arbitrary integration constants. It is noteworthy that the Weyl transformation \cite{Carmeli}
\begin{equation}\label{Weyl}
\Phi(\rho,z)=e^{-i\alpha}\coth{\Psi}
\end{equation}
is a special case of (\ref{gen_f}) when the plus sign is chosen and bith constants are set as $c_1=1=c_2$. Employing both (\ref{Weyl}) and (\ref{Ernst}), it is not difficult to verify that 
\begin{equation}
f=\frac{1}{2\cosh^2{\Psi}+2\cos{\alpha}\sinh{\Psi}\cosh{\Psi}-1},\quad
u=\frac{2\sin{\alpha}\sinh{\Psi}\cosh{\Psi}}{1-2\cosh^2{\Psi}-2\cos{\alpha}\sinh{\Psi}\cosh{\Psi}}.
\end{equation}
Consequently, the governing equations for $w$ and $\mu$ become
\begin{equation}
\partial_\rho w=-2\rho\sin{\alpha}\partial_z\Psi,\quad
\partial_z w=2\rho\sin{\alpha}\partial_\rho\Psi
\end{equation}
and
\begin{equation}\label{mu}
\partial_\rho(\mu+\ln{f})=2\rho\left[(\partial_\rho\Psi)^2-(\partial_z\Psi)^2\right],\quad
\partial_z(\mu+\ln{f})=4\rho\partial_\rho\Psi\partial_z\Psi.
\end{equation}
At this point, a brief comment is in order. First of all, \cite{GP} derived equations similar to (\ref{mu}) where $\gamma=(\mu+\ln{f})/2$ and $U\equiv\Psi$. Nonetheless, there is a typographical error in \cite{GP} regarding the first equation in $(10.4)$: the plus sign should be substituted with a minus sign. As highlighted by \cite{Carmeli}, only the solutions with $\alpha=0$ have physical relevance. For this scenario, we obtain
\begin{equation}\label{fpsi}
f=\frac{1-\tanh{\Psi}}{1+\tanh{\Psi}}=e^{-2\Psi},\quad
u=0,\quad
\partial_\rho w=0=\partial_z w.
\end{equation}
The equations in (\ref{mu}) remain unchanged. It is evident that for the line element (\ref{Weyl-Lewis-Papa}) to asymptotically approach the Minkowski metric, the condition $w\equiv 0$ must hold together with $\Psi\to 0$ and $e^\mu\to 1$ as $\rho,z\to\infty$. Finally, we recall that in the Newtonian limit, the metric tensor can be approximated as $g_{\alpha\beta}=\eta_{\alpha\beta}+h_{\alpha\beta}$, where $\eta_{\alpha\beta}$ denotes the Minkowski metric tensor, $h_{\alpha\beta}$ is a small correction and
\begin{equation}\label{gravpot}
g_{00}=1-2\Psi+\mathcal{O}(\Psi^2).
\end{equation}
As indicated by \cite{Carmeli,GP}, a common approach to constructing a cylindrically symmetric solution begins with selecting an exact Newtonian/Coulomb potential $\Psi$ for some axially symmetric physical system in a flat space described by standard cylindrical coordinates. Then, the function $f$ is derived from (\ref{fpsi}), and $\mu$ is determined  by solving the system (\ref{mu}). Subsequently, the solution is interpreted as the gravitational field corresponding to the Newtonian source. Nevertheless, \cite{GP} pointed out that this method might not always yield the appropriate physical interpretation of the derived line element. A possible explanation put forward by \cite{Futa,coop1,coop2,Carr, Caccia} is that the Newtonian approximation is locally applicable everywhere for slow and weak gravitational fields, but not globally. Even in cases with low energy density and particle velocities, General Relativity can encompass non-Newtonian phenomena, including propagating gravitational waves \cite{gw}, gravitational shielding \cite{Carlotto}, and stationary vacuum solutions, known as geons \cite{Anderson}. To circumvent the challenges inherent in the above-described method, we decided to follow a different strategy in the next two sections, relying on the use of the Hankel transform.

\section{Metrics generated by the Hankel transform}
The axisymmetric Laplace equation in cylindrical coordinates $(\rho,\varphi,z)$ for the unknown function $\Psi$ reads
\begin{equation}\label{Lap}
\frac{1}{\rho}\partial_\rho\left(\rho\partial_\rho\Psi\right)+\partial_{zz}\Psi=0,\quad
\Psi=\Psi(\rho,z),\quad
0<\rho<\infty,\quad z>0.
\end{equation}
As we will see, the condition $z>0$ is not too restrictive because one can still construct solutions to the equation above having the property of vanishing as $z\to\pm\infty$. We are interested in solving (\ref{Lap}) subject to the following boundary conditions
\begin{enumerate}
\item
$\Psi\to 0$ as $\rho$ and $z\to\infty$;
\item
any additional condition ensuring that the metric becomes Minkowski asymptotically at space-like infinity.
\end{enumerate}
Since the problem is axisymmetric, it is convenient to introduce the zero order Hankel transform \cite{Deb} which is defined as follows
\begin{equation}
\mathcal{H}_0\left\{f(\rho)\right\}=\widehat{f}(k)=\int_0^\infty\rho J_0(k\rho)f(\rho)~d\rho,
\end{equation}
where $f$ is a suitable function and $J_0$ denotes the zero order Bessel function of the first kind. The zero order inverse Hankel transform is
\begin{equation}
\mathcal{H}_0^{-1}\left\{\widehat{f}(k)\right\}=f(\rho)=\int_0^\infty k J_0(k\rho)\widehat{f}(k)~dk.
\end{equation}
If we apply $\mathcal{H}_0$ to (\ref{Lap}) together with $7.3.12$ in \cite{Deb}, we obtain
\begin{equation}
\partial_{zz}\widehat{\Psi}-k^2\widehat{\Psi}=0,\quad
\widehat{\Psi}=\widehat{\Psi}(k,z)
\end{equation}
whose general solutions is
\begin{equation}\label{HT}
\widehat{\Psi}(k,z)=A(k)e^{-kz}+B(k)e^{kz}.
\end{equation}
The first boundary condition requires that $B(k)\equiv 0$ while $A(k)$ is fixed by the second boundary condition. Hankel transforming back (\ref{HT}) gives the following integral representation for the solution to (\ref{Lap}), namely
\begin{equation}\label{sol_int}
\Psi(\rho,z)=\int_0^\infty kJ_0(k\rho)A(k)e^{-kz}~dk.
\end{equation}
As a side note, we observe that if we relax the boundary conditions above by requiring that $\Psi$ vanishes only for $\rho\to\infty$, it is possible to construct a solution of the Laplace equation exhibiting an oscillatory behaviour in the $z$-direction. An example is provided by the problem
\begin{equation}\label{LapI}
\frac{1}{\rho}\partial_\rho\left(\rho\partial_\rho\Psi\right)+\partial_{zz}\Psi=0,\quad
\Psi=\Psi(\rho,z),\quad
0<\rho<\infty,\quad -\infty<z<+\infty
\end{equation}
together with the mixed boundary data
\begin{equation}
\lim_{\rho\to 0}\rho^2\Psi(\rho,z)=0,\quad
\lim_{\rho\to 0}\rho\partial_\rho\Psi=-Af(z)\quad\mbox{on}\quad  -\infty<z<+\infty
\end{equation}
with some positive constant $A$ and some suitable function $f(z)$. Then, according to \cite{Deb} one finds
\begin{equation}
\widehat{\Psi}(k,z)=\frac{A}{k}\int_{-\infty}^{+\infty}e^{-k|z-\xi|}f(\xi)d\xi
\end{equation}
and the corresponding solution of the mixed boundary value problem is
\begin{equation}
\Psi(\rho,z)=A\int_{-\infty}^{+\infty}\frac{f(\xi)}{\sqrt{\rho^2+(z-\xi)^2}}d\xi.
\end{equation}
Let $z-\xi=\zeta$. Then, the integral representation for $\Psi$ becomes
\begin{equation}
\Psi(\rho,z)=A\int_{-\infty}^{+\infty}\frac{f(z-\zeta)}{\sqrt{\rho^2+\zeta^2}}d\zeta.
\end{equation}
Let $\widehat{\alpha}$ be a positive real parameter. If we choose $f(z)=\sin{(\widehat{\alpha} z)}$, realize that $\sin{(\widehat{\alpha} z)}/\sqrt{\rho^2+\zeta^2}$ is an odd function and apply $3.754.2$ in \cite{Grad}, we find that
\begin{equation}
\Psi(\rho,z)=2AK_0(\widehat{\alpha}\rho)\sin{(\widehat{\alpha} z)},
\end{equation}
where $K_0$ denotes the zero order modified Bessel function of the second kind. We recall that $K_0$ decays exponentially as $\rho\to\infty$ while it displays a logarithmic divergence for $\rho\to 0$. According to (\ref{fpsi}), the metric coefficient $f$ is
\begin{equation}\label{g00geon}
g_{00}=f=e^{-4AK_0(\widehat{\alpha}\rho)\sin{(\widehat{\alpha} z)}}.
\end{equation}
Note that $g_{00}$ admits the following asymptotic expansion in $\rho$ for fixed $z$
\begin{equation}
g_{00}=1-2A\sqrt{\frac{2\pi}{\widehat{\alpha}\rho}}e^{-\widehat{\alpha}\rho}\sin{(\widehat{\alpha}z)}+\mathcal{O}\left(\frac{e^{-2\widehat{\alpha}\rho}}{\rho}\right)
\end{equation}
from which we can evince that $g_{00}\to 1$ as $\rho\to\infty$. Furthermore, it is straightforward to verify that $g_{00}\equiv 1$ on the equatorial plane $z=0$. Concerning the behaviour of $g_{00}$ for $\rho\to 0$ while $z$ is kept fixed, the following expansion holds
\begin{equation}
g_{00}=\left(\frac{\widehat{\alpha}\rho}{2}\right)^{4A\sin{(\widehat{\alpha}z)}}e^{4A\gamma\sin{(\widehat{\alpha}z)}}\left[1+\mathcal{O}(\rho^2)\right],
\end{equation}
where $\gamma$ is the Euler-Mascheroni constant. Since both $A$ and $\widehat{\alpha}$ are positive, we immediately see that $g_{00}$ becomes singular on $\rho=0$ whenever $\sin{(\widehat{\alpha}z)}<0$. More precisely, we observe that such a divergent behaviour occurs for $\rho\to 0$ only when  
\begin{equation}
\frac{\pi}{\widehat{\alpha}}(1+2m)<z<\frac{2\pi}{\widehat{\alpha}}(1+m),\quad m\in\mathbb{Z}.
\end{equation}
In other words, $g_{00}$ displays a periodic singular behaviour along the $z$-axis. 

Specifically, on the plane $z=3\pi/2$, we find that $g_{00}\to\infty$ as $\rho\to 0$. This leads to a central redshift, $Z=1/\sqrt{g_{00}} - 1$, approaching $-1$. This is a highly counter-intuitive result as it implies an extreme blueshift, seemingly suggesting that the light source moves at the speed of light towards the observer. However, in this case, we are dealing with a stationary source containing a naked singularity, a condition where conventional rules of spacetime may not fully apply and the highly warped spacetime, might produce such an intense gravitational field that an extreme blueshift may occur. While speculative, such phenomena might indeed occur in the vicinity of naked singularities as they have a profound effect on the surrounding spacetime fabric. In fact, the occurrence of negative redshift is not solely exclusive to our scenario but has been also reported in the context of certain wormhole solutions \cite{SHK}. Moreover, note that $g_{00}$ is instead regular whenever $\sin{(\widehat{\alpha}z)}\geq 0$ as it can be seen in Fig.~\ref{figure1geon}.
\begin{figure}[!ht]\label{unz}
\centering
    \includegraphics[width=0.4\textwidth]{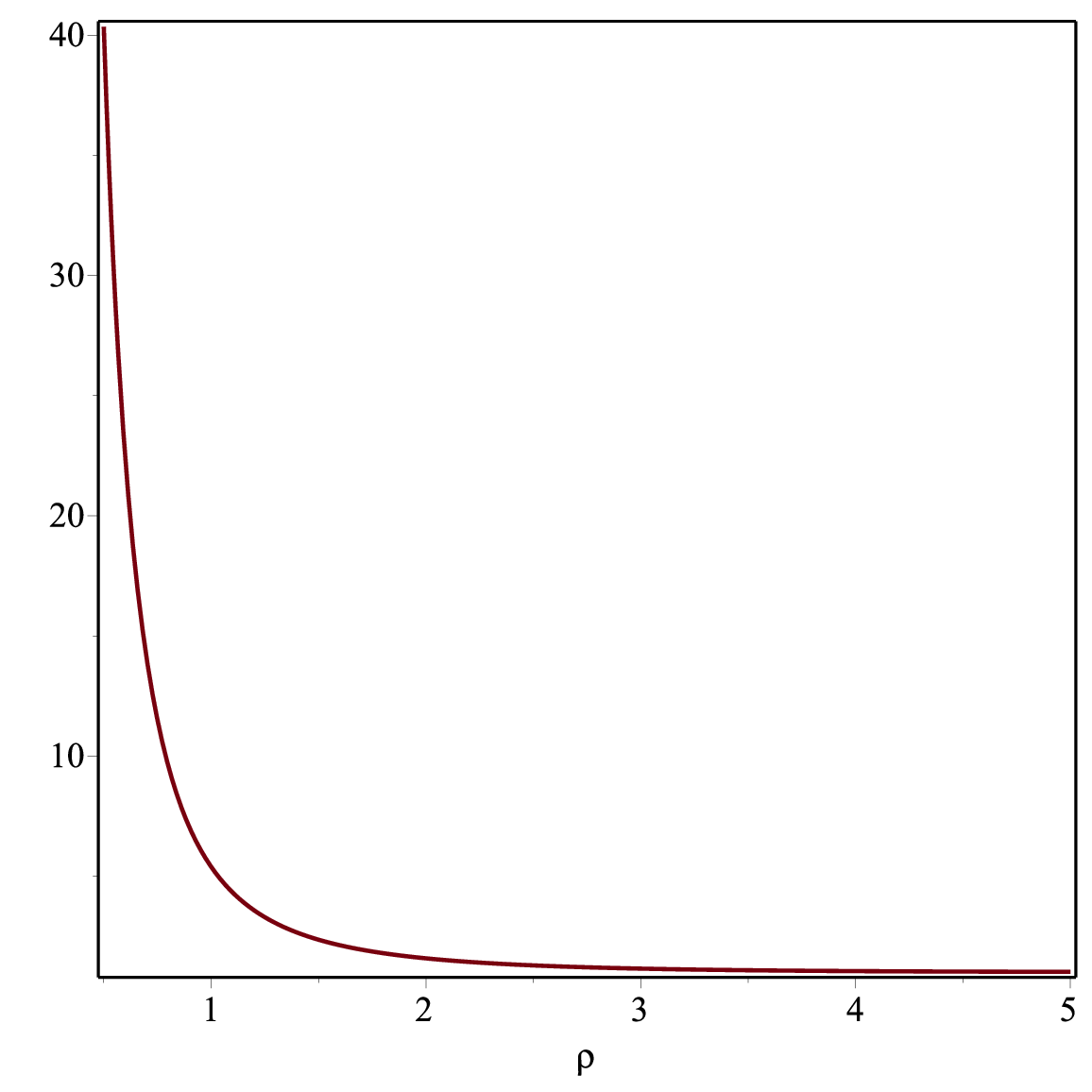}
    \includegraphics[width=0.4\textwidth]{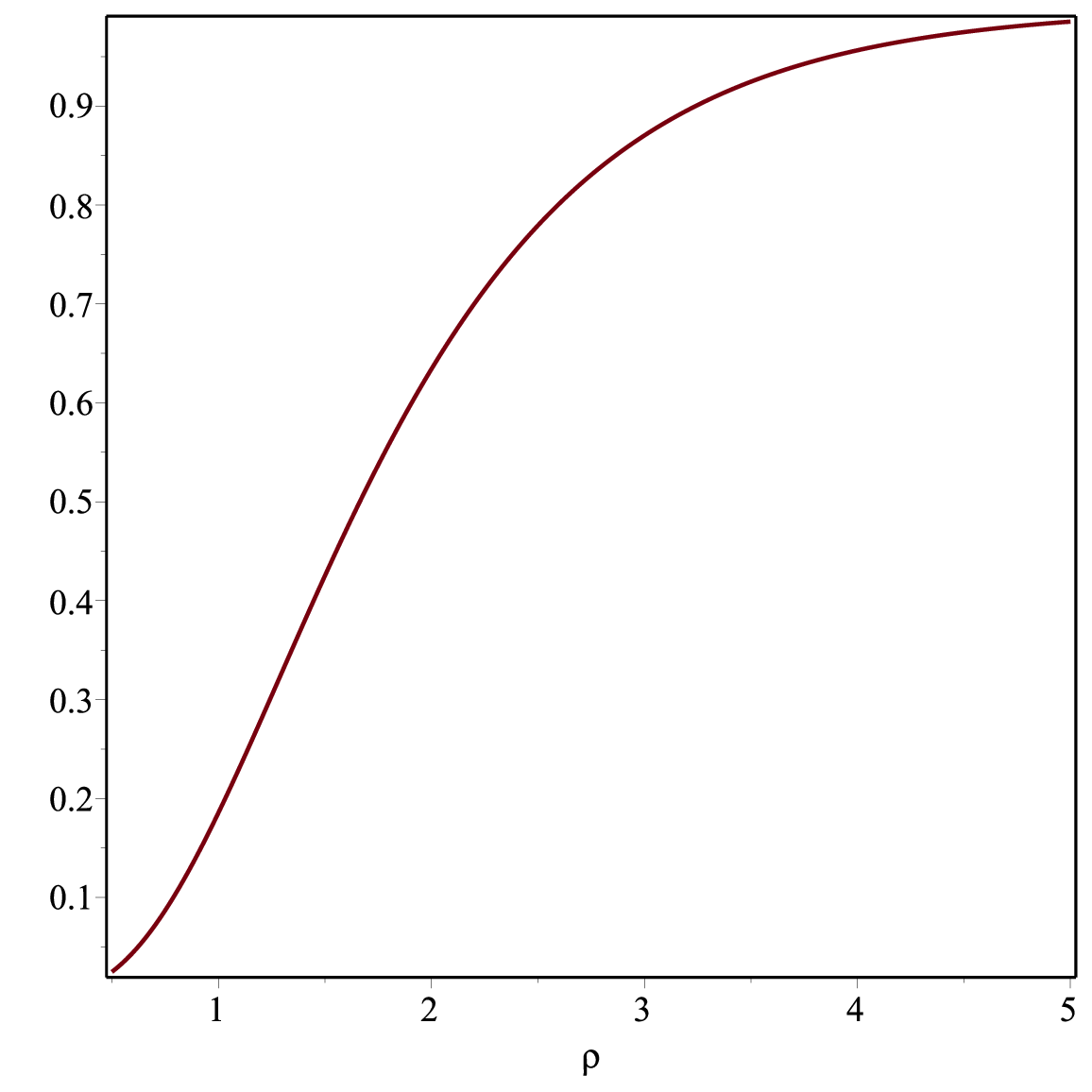}
\caption{\label{figure1geon}
Plots of the metric coefficient $g_{00}$ given in (\ref{g00geon}) for $A=\widehat{\alpha}=1$. The figure on the left describes a divergent behaviour for  $g_{00}$ on the plane $z=3\pi/2$ as $\rho\to 0$ while the figure on the right shows that $g_{00}$ is regular on the plane $z=\pi/2$ and vanishes in the aforementioned limit.}
\end{figure}
Even though $g_{00}$ can never vanish on the equatorial plane, we observe that $g_{00}=0$ at $\rho=0$ for every $z\in(2m\pi/\widehat{\alpha},\pi(1+2m)/\widehat{\alpha})$. In order to discuss the nature of the singularities appearing in $g_{00}$, it is necessary to look into the Kretschmann invariant $K=R_{\alpha\beta\gamma\delta}R^{\alpha\beta\gamma\delta}$. To this purpose, we need to obtain the metric coefficient $e^\mu$. In that regard, integrating the second equation in (\ref{mu}) leads to
\begin{equation}\label{kon1}
\mu+\ln{f}=-2A^2\alpha\rho K_0(\widehat{\alpha}\rho)K_1(\widehat{\alpha}\rho)\sin^2{(\widehat{\alpha} z)}+H(\rho),
\end{equation}
where $H(\rho)$ is an unknown function satisfying the first order differential equation 
\begin{equation}\label{ODEH}
\frac{dH}{d\rho}+2A^2\widehat{\alpha}^2\rho K_0^2(\widehat{\alpha}\rho)=0.
\end{equation}
The latter equation has been obtained by substituting (\ref{kon1}) into the first equation in (\ref{mu}). Integrating (\ref{ODEH}) with Maple yields
\begin{equation}
H(\rho)=c_1+A^2\widehat{\alpha}^2\rho^2\left[K_1^2(\widehat{\alpha}\rho)-K_0^2(\widehat{\alpha}\rho)\right]
\end{equation}
with $c_1$ an arbitrary integration constant which must be chosen to be zero so that the line element (\ref{Weyl-Lewis-Papa}) goes over into the Minkowski metric as $\rho\to\infty$. Let
\begin{equation}
F\equiv F(\rho,z)=A^2\widehat{\alpha}^2\rho^2\left[K_1^2(\widehat{\alpha}\rho)-K_0^2(\widehat{\alpha}\rho)\right]-2A^2\widehat{\alpha}\rho K_0(\widehat{\alpha}\rho)K_1(\widehat{\alpha}\rho)\sin^2{(\widehat{\alpha} z)}.
\end{equation}
Then, it is straightforward to check that the metric coefficients $g_{\rho\rho}$ and $g_{zz}$ are given by
\begin{equation}
g_{\rho\rho}=g_{zz}=e^{2\Psi+F}.
\end{equation}
Moreover, for fixed $z$ and $\rho\to\infty$
\begin{equation}
e^{2\Psi+F}=1+2A\sqrt{\frac{2\pi}{\widehat{\alpha}\rho}}e^{-\widehat{\alpha}\rho}\sin{(\widehat{\alpha}z)}+\mathcal{O}\left(e^{-2\widehat{\alpha}\rho}\right)
\end{equation}
thus signalizing that in this regime both $g_{\rho\rho}$ and $g_{zz}\to 1$. Finally, in order to understand whether the metric coefficient $g_{00}$ is plagued by coordinate or curvature singularities, we computed the Kretschmann scalar with Maple. Since the corresponding analytic expression for $K$ is extremely lengthy, we limit us here to exhibit $K$ on the equatorial plane, namely
\begin{equation}
\left.K\right|_{z=0}=4\widehat{\alpha}^2 A^2 e^{2A^2\widehat{\alpha}^2\rho^2\left[K_0^2(\widehat{\alpha}\rho)-K_1^2(\widehat{\alpha}\rho)\right]}\left[16\left(\widehat{\alpha}K_1(\widehat{\alpha}\rho)-A^2\widehat{\alpha}^2K_0^3(\widehat{\alpha}\rho)\right)^2+123A^2\widehat{\alpha}^2 K_0^4(\widehat{\alpha}\rho)\right],
\end{equation}
while the behaviour of $K$ for different values of $z$ has been displayed in Fig.~\ref{figure2sin} from which we observe that the metric exhibits a curvature singularity along the whole $z$-axis.
\begin{figure}[!ht]\label{bunz}
\centering
    \includegraphics[width=0.3\textwidth]{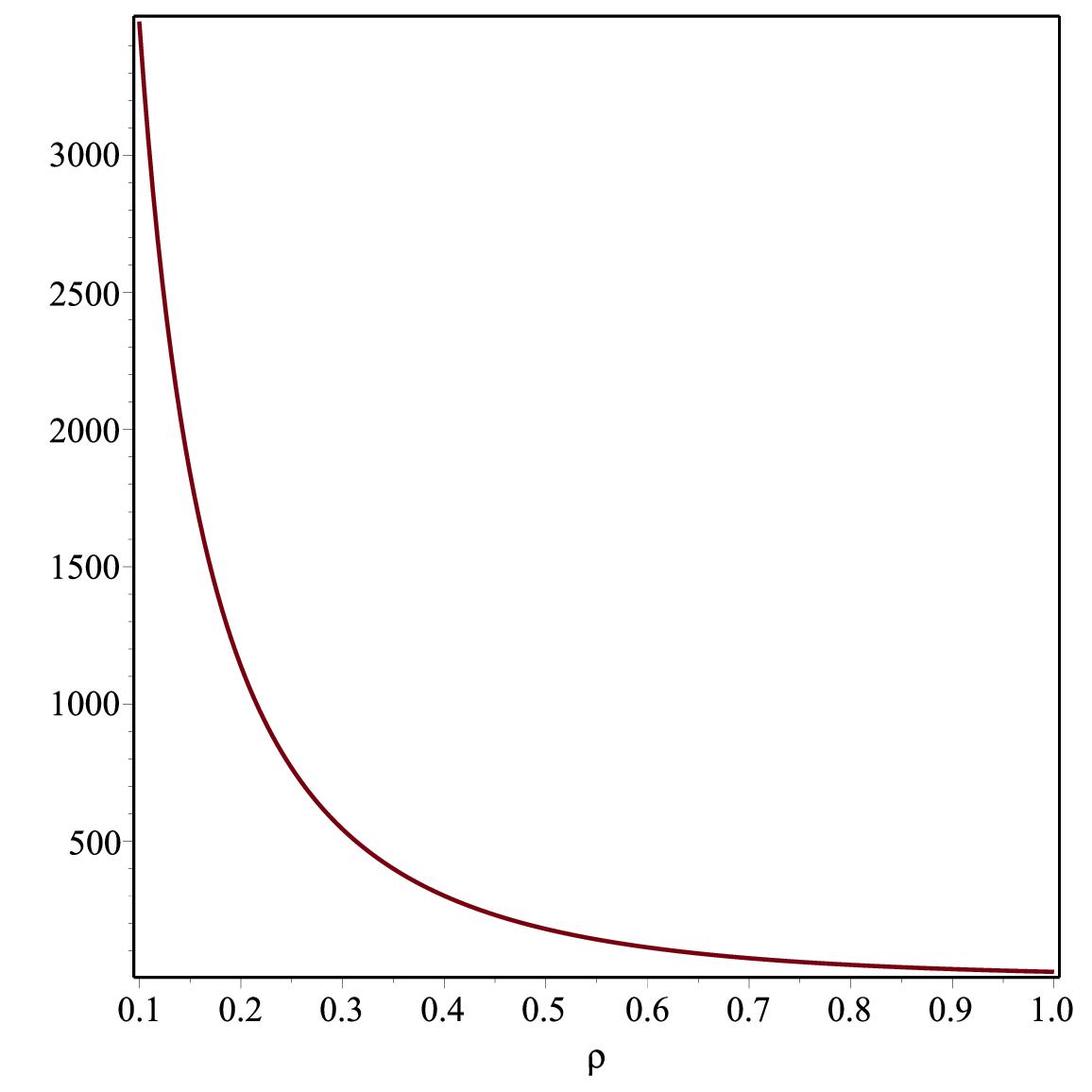}
    \includegraphics[width=0.3\textwidth]{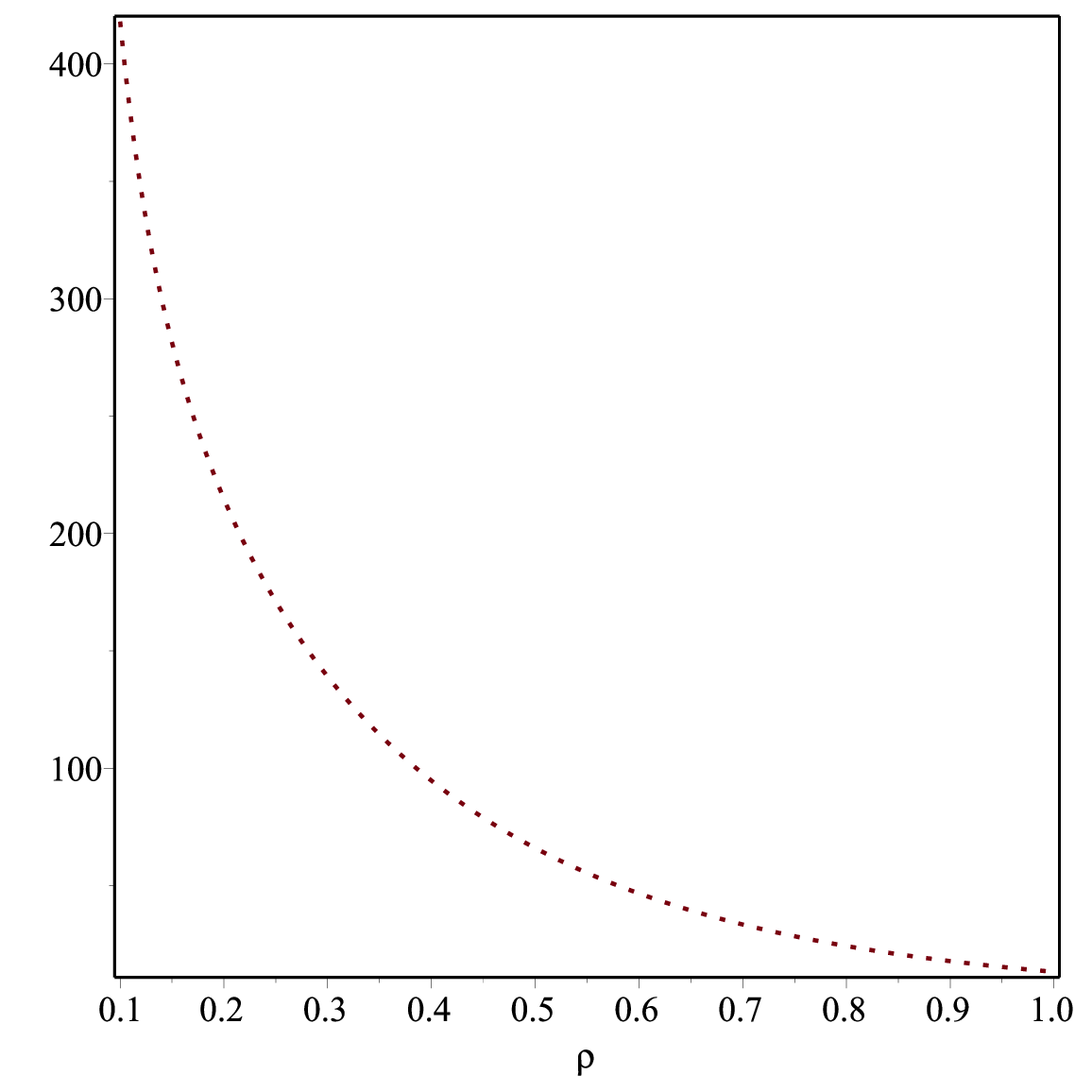}
    \includegraphics[width=0.3\textwidth]{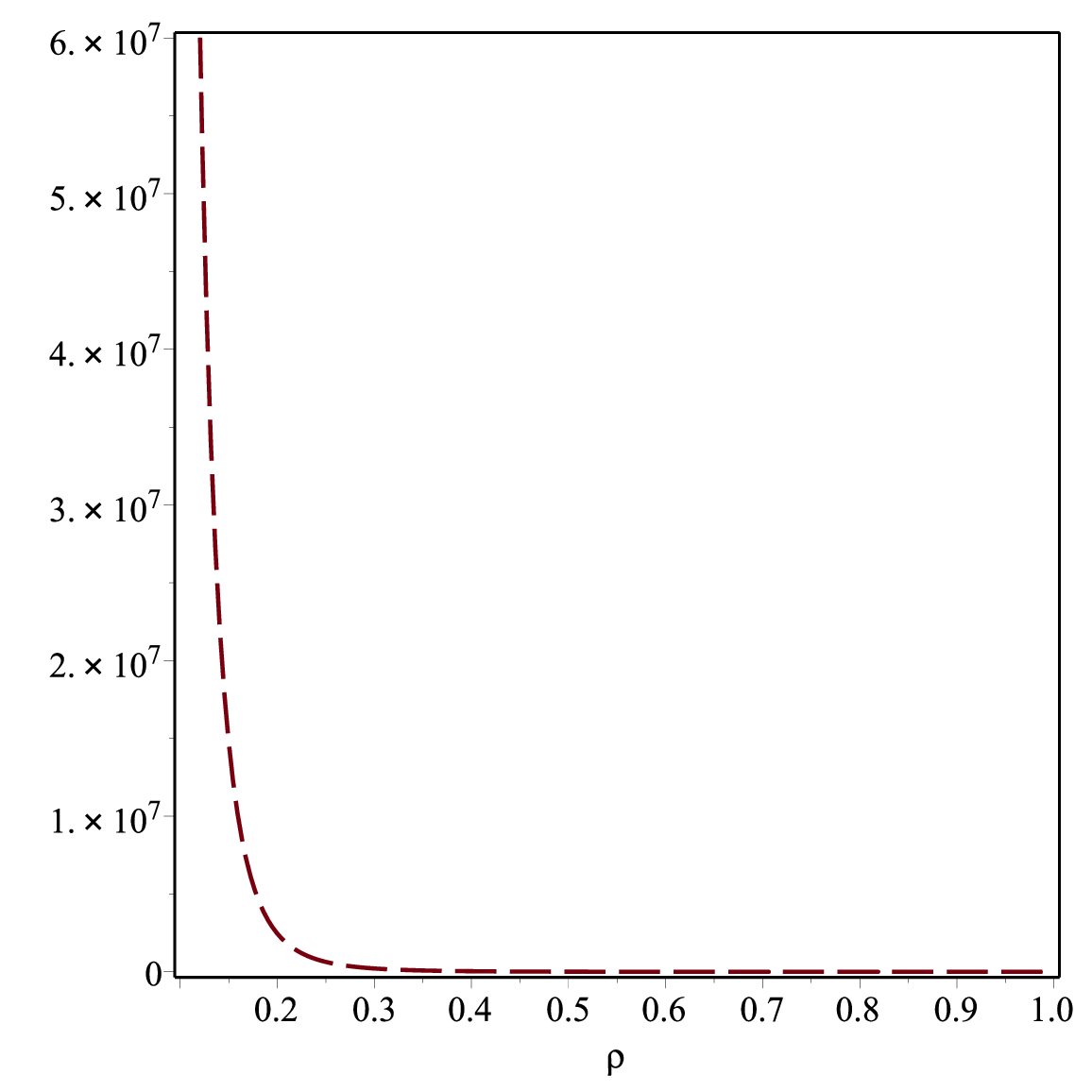}
\caption{\label{figure2sin}
Plots of the Kretschmann scalar $K$ for $A=\widehat{\alpha}=1$. The figures describe the divergent behaviour of $K$ at $\rho=0$ for the planes $z=0$ (solid line, left panel), $z=0.1$ (dotted line, middle panel) and $z=3.5$ (space-dotted line, right panel). Note that for $z=\pi$ the corresponding plot is again given by that for $z=0$ due to the fact that the metric coefficients depend on the periodic function $\sin{(\widehat{\alpha}z)}$.}
\end{figure}

\subsection{The Curzon solution}
We present an alternative method based on the use of the Hankel transform which allows to derive the Curzon metric \cite{Curzon}. It is not difficult to verify that such a  metric can be obtained by a certain limiting process from the boundary value problem
\begin{eqnarray}
\frac{1}{\rho}\partial_\rho\left(\rho\partial_\rho\Psi\right)+\partial_{zz}\Psi&=&0\quad
\mbox{on}\quad
0<\rho<\infty,\quad z>0,\\
\Psi(\rho,0)&=&\frac{\Psi_0}{\sqrt{a^2+\rho^2}},\quad a>0,~0<\rho<\infty,\quad z>0,\label{C1}\\
\Psi(\rho,z)&\to&0\quad\mbox{as}\quad z\to+\infty~\forall\rho>0.
\end{eqnarray}
The Hankel transform of (\ref{C1}) can be easily computed with Maple and is found to be
\begin{equation}
\mathcal{H}_0\left\{\Psi(\rho,0)\right\}=\Psi_0\frac{e^{-ak}}{k}.
\end{equation}
This information gives the $A(k)$ we need to replace in (\ref{sol_int}). Hence, we end up with the following integral representation
\begin{equation}
\Psi(\rho,z)=\Psi_0\int_0^\infty e^{-k(z+a)}J_0(\rho k)~dk.
\end{equation}
Using $6.611.1$ in \cite{Grad}, i.e.
\begin{equation}
\int_0^\infty e^{-\gamma x}J_\nu(\beta x)~dx=\frac{\beta^{-\nu}\left(\sqrt{\gamma^2+\beta^2}-\gamma\right)^\nu}{\sqrt{\gamma^2+\beta^2}},\quad
\Re(\nu)>-1,\quad\Re{(\gamma+i\beta)}>0
\end{equation}
with $\gamma=z+a$, $\beta=\rho$ and $\nu=0$ yields
\begin{equation}
\Psi(\rho,z)=\frac{\Psi_0}{\sqrt{(z+a)^2+\rho^2}}.
\end{equation}
Note that the Curzon solution is recovered in the limit $a\to 0$. As a final remark, we would like to observe that the limiting process and the boundary data needed to reproduce the Curzon metric via Hankel transform are not unique. We can convince ourselves that this is the case by considering the following Neumann problem
\begin{eqnarray}
\frac{1}{\rho}\partial_\rho\left(\rho\partial_\rho\Psi\right)+\partial_{zz}\Psi&=&0\quad
\mbox{on}\quad
0<\rho<\infty,\quad z>0,\\
\left.\partial_z\Psi(\rho,z)\right|_{z=0}&=&-\frac{2\Psi_0}{a^2}H(a-\rho),\quad \mbox{for}~0<\rho<\infty,\\
\Psi(\rho,z)&\to&0\quad\mbox{as}\quad z\to+\infty~\forall\rho>0,
\end{eqnarray}
where $H$ denotes the Heaviside function. It is not difficult to check that in the limit of $a\to 0$, the solution of the above problem reproduces the Curzon solution, i.e.
\begin{equation}
\lim_{a\to 0}\Psi(\rho,z)=\frac{\Psi_0}{\sqrt{\rho^2+z^2}}.
\end{equation}
To this purpose, we recall that the solution of the Laplace equation with boundary conditions as above is \cite{Deb}
\begin{equation}
\Psi(\rho,z)=\frac{2\Psi_0}{a}\int_0^\infty\frac{1}{k}J_1(ak)J_0(k\rho)e^{-kz}dk,
\end{equation}
which is a special case of the integral 
\begin{equation}
I(\mu,\nu;\lambda)=\int_0^\infty e^{-pt}t^\lambda J_\mu(\widetilde{a}t)J_\nu(\widetilde{b}t)dt
\end{equation}
studied on page $314$ in \cite{Luke}. However, the solution of such an integral results in an extremely complicated combination of elliptic functions. Even though it allows to compute the metric function $f$ in a relatively straightforward way, it makes the computation of $\mu$ by quadratures from (\ref{mu}) a formidable task. By means of the Lebesgue Dominated Convergence Theorem and taking into account that $J_1(ak)/a\to k/2$ as $a\to 0$, it follows that
\begin{equation}\label{2.32}
\lim_{a\to 0}\Psi(\rho,z)=\Psi_0\int_0^\infty J_0(k\rho)e^{-kz}dk=\frac{\Psi_0}{\sqrt{\rho^2+z^2}},
\end{equation}
where the last integral has been evaluated with Maple. We conclude this part by offering a simple mathematical argument which not only differs from those existing in the literature but also sheds some light on the nature and complexity of the singularity at $\rho=0=z$. To this purpose, it is useful to recall that \cite{Gaut} was the first to observe that the Kretschmann scalar may or may not blow up as $R=\sqrt{\rho^2+z^2}\to 0$ depending on which direction is chosen to approach the singular point $(\rho,z)=(0,0)$. On the other hand, \cite{Stac} focussed on the size of such a singularity. By switching to spherical coordinates $(R,\vartheta,\varphi)$, the author considered the area of the surface $t=const$ and $g_{00}=const$. In particular, he showed that the area of the gravitational equipotential surfaces gets smaller and smaller as $R$ decreases from infinity until it exhibits a minimum. However, as one allows $R$ to further decrease, the area increases without bound as $R\to 0^+$. A further refinement of the work in \cite{Gaut} is represented by \cite{CJ} where the authors came to the conclusion that instead of talking of a directional singularity at $R=0$, it would be more appropriate to refer to it as a trajectory singularity. Ref. \cite{SM}, instead,  adopted a different perspective. More precisely, the starting point there is the observation that the regular behaviour of the Kretschmann scalar along the axis $\rho=0$ despite its divergent behaviour for all other directions of approach to $R=0$ might hint to the fact that test particles travelling to $R=0$ along $\rho=0$ could get access to some new region. By considering null geodesics on a fixed plane $\varphi=const$ and introducing comoving coordinates, they showed that the point-like appearance of $R=0$ is quite tricky and one should think of it as an infinite plane ($z=0$) at which the space-time becomes flat for each slice $t=const$. Finally, the authors in \cite{SS} were able to set up a compactified coordinate chart for the hypersurface $t=const$ allowing to show that the singularity at $R=0$ appears as a ring such that space-like geodesics can hit it in finite proper distance. Moreover, they not only showed that such a ring displays the counter-intuitive property of having finite radius while displaying an infinite circumference but they also found that the manifold exhibits a double-sheeted topology inside the ring. At this point, we would like to point out that the complexity of the singularity at $R=0$ already emerges from the following simple observation. First of all, even without computing the integral appearing in (\ref{2.32}) the potential function $\Psi$ (see equation (\ref{gravpot})) is expected to become singular at the origin $(\rho,z)=(0,0)$ because one integrates a constant function on an interval of infinite length. Moreover, since the integration of the restriction of the function in (\ref{2.32}) on the plane $z=0$ gives the result $1/\rho$, $\Psi$ is usually interpreted as the Newtonian potential of a point-like unit mass at the origin. This argument commonly used in the literature (see for instance \cite{Carmeli}) should be taken with some caution because if we compute the same integral in (\ref{2.32}) by approaching $z=0$ along the line $z=\rho$, then instead of getting $1/\rho$ we end up with a different Newtonian potential, namely $1/(\sqrt{2}\rho)$. This sensitivity on the direction along which the origin is approached seems to suggest that the potential arising from the integral in (\ref{2.32}) might have a complicated essential singularity at the origin.

\subsection{The arcsine metric}
We show that it is possible to construct a nontrivial metric which is not plagued by a naked singularity as is the case for the Curzon metric and moreover, it goes over to the Minkowsky metric as  $\rho,z\to\infty$. An additional interesting feature of our solution is that it turns out to be symmetric under reflection with respect to the plane $z=0$. Such a reflection symmetric solution to the Ernst equation might be physically relevant because as already pointed out by \cite{Lind,Meinel} reflection symmetry is a key ingredient for a very large class of equilibrium stellar models. Let us consider the following boundary value problem inspired by a similar one in electrostatics concerning an electrified disk of radius $R>0$ in the plane $z=0$ and centred at the origin, namely
\begin{eqnarray}
&&\frac{1}{\rho}\partial_\rho\left(\rho\partial_\rho\Psi\right)+\partial_{zz}\Psi=0\quad
\mbox{on}\quad
0<\rho<\infty,\quad 0<z<\infty,\\
&&\Psi(\rho,0)=\Psi_0\quad\mbox{on}\quad 0\leq\rho<R,\\
&&\left.\partial_z\Psi(\rho,z)\right|_{z=0}=0\quad\mbox{on}\quad R<\rho<\infty\quad\mbox{and}\quad\Psi\to 0\quad\mbox{as}\quad z\to\infty~\forall\rho\geq 0.\label{bclast}
\end{eqnarray}
If we insist to interpret $\Psi$ as the Newtonian potential of some massive source, then according to the boundary data prescribed above, such a source should be seen as an infinitesimally thin disk of radius $R$ while the condition (\ref{bclast}) simply states that the gravitational force acting on a test particle on the equatorial plane is purely radial. Proceeding as in \cite{Deb}, it can be shown that the solution is
\begin{equation}
\Psi(\rho,z)=\frac{2\Psi_0}{\pi}\int_0^\infty J_0(k\rho)\frac{\sin{(Rk)}}{k}e^{-kz}dk.
\end{equation}
The above integral can be computed by means of $6.752.1$ in \cite{Grad}, i.e.
\begin{equation}
\int_0^\infty J_0(bx)\frac{\sin{(cx)}}{x}e^{-ax}dx=\arcsin{\left(\frac{2c}{\sqrt{a^2+(c+b)^2}+\sqrt{a^2+(c-b)^2}}\right)}
\end{equation}
subject to the conditions $\Re{(a)}>|\Im{(b)}|$ and $c>0$. In the present case, $a=z$, $c=R$, $b=\rho$ so $\Im{(b)}=\Im{(\rho)}=0$ and the constraint $\Re{(a)}>|\Im{(b)}|$ is just the condition $z>0$. Let 
\begin{equation}
\Delta_\pm=(\rho\pm R)^2+z^2.
\end{equation}
Then, we find
\begin{equation}\label{psi}
\Psi(\rho,z)=\frac{2\Psi_0}{\pi}\arcsin{\left(\frac{2R}{\sqrt{\Delta_+}+\sqrt{\Delta_{-}}}\right)}=\frac{2\Psi_0}{\pi}\arcsin{\left(\frac{\sqrt{\Delta_+}-\sqrt{\Delta_{-}}}{2\rho}\right)}.
\end{equation}
Note that asymptotically away $\rho^2+z^2\approx r^2$ with $\rho\approx r\sin{\vartheta}$ and in that regime
\begin{equation}\label{apross}
g_{00}=e^{-2\Psi}=1-\frac{4\Psi_0 R}{\pi r}+\mathcal{O}\left(\frac{1}{r^2}\right)
\end{equation}
from which we conclude that $4\Psi_0/\pi=2M/R$ where $M$ is the total mass of the gravitational object. Hence, the metric coefficient $g_{00}$ turns out to be
\begin{equation}\label{g00arcsin}
g_{00}=f=\mbox{exp}\left(-\frac{M}{R}\arcsin{\left(\frac{2R}{\sqrt{\Delta_+}+\sqrt{\Delta_{-}}}\right)}\right).
\end{equation}
It is gratifying to observe that $g_{00}\to 1$ at space-like infinity. At this point some comments are in order. First of all, as a double check we verified with Maple that the above solution satisfies the Laplace equation. Moreover, a trivial computation shows that
\begin{equation}
\Psi(\rho,0)=\left\{
\begin{array}{cc}
\Psi_0 & \mbox{if}~0\leq\rho<R,\\
\frac{2\Psi_0}{\pi}\arcsin{\left(\frac{R}{\rho}\right)}&\mbox{if}~\rho\geq R.
\end{array}
\right.
\end{equation}
This signalizes that $\Psi$ is continuous on $z=0$ and $\rho=R$ and is clearly continuous elsewhere. Furthermore, using the first representation for $\Psi$ in (\ref{psi}) yields
\begin{equation}
\partial_z\Psi=\frac{Mz}{\pi\sqrt{\Delta_+\Delta_{-}}}\frac{\sqrt{\Delta_+}-\sqrt{\Delta_{-}}}{\sqrt{\Delta_+}+\sqrt{\Delta_{-}}}\frac{1}{\sqrt{(\sqrt{\Delta_+}+\sqrt{\Delta_{-}})^2-4R^2}}.
\end{equation}
At this point, it is trivial to check that the condition $\left.\partial_z\Psi(\rho,z)\right|_{z=0}=0$ is indeed fulfilled for $\rho>R$. Additional information about the metric coefficient $f$ can be gained from the inspection of its plot. To this purpose, it is convenient to introduce the rescaled variables $u=\rho/R$ and $v=z/R$. As it can be seen from Figure~3, $g_{00}$ exhibits a cusp singularity along the ring $\rho=R$ located on the plane $z=0$. To understand whether this is a curvature or a coordinate singularity, it is necessary to analyse the Kretschmann scalar. To this purpose, we now derive the remaining metric coefficient $e^\mu$ entering in (\ref{Lewis-Papa}).
\begin{figure}[!ht]\label{uno}
\centering
    \includegraphics[width=0.5\textwidth]{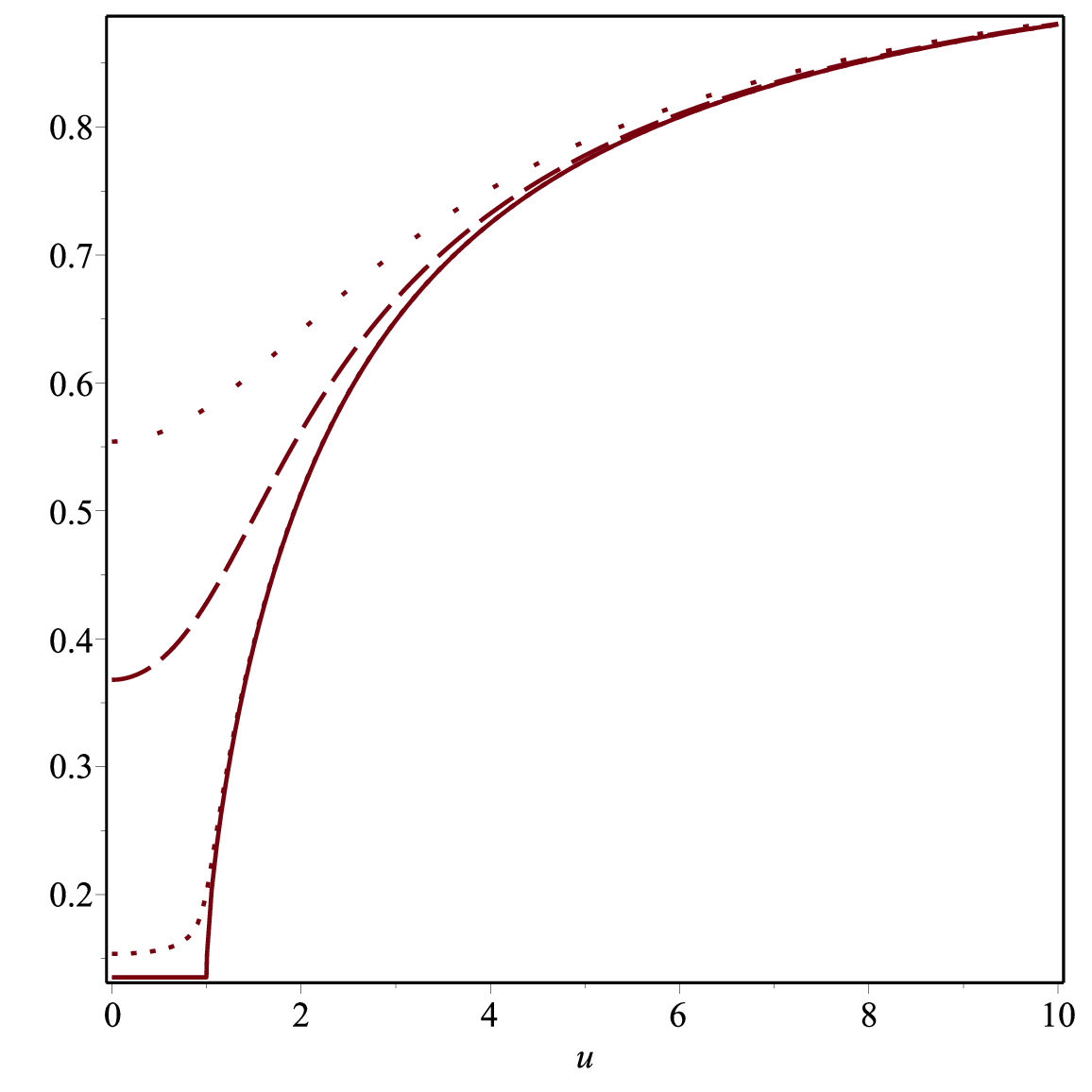}
\caption{\label{figure1}
Plot of the metric coefficient $g_{00}=f$ defined in (\ref{g00arcsin}) as a function of $u=\rho/R$ for different values of $v=z/R$ in the case $R=M$. The solid, dotted, dashed and space-dotted lines correspond to $v=0$, $v=0.1$, $v=1$ and $v=2$, respectively. Moreover, $f$ does not vanish at the origin but it has the value $f(0,0)=0.13533$.}
\end{figure}
First of all, we  observe that the second equation in (\ref{mu}) can be integrated with Maple. In particular, we find that
\begin{equation}\label{ascia}
\mu+\ln{f}=T(\rho,z)+H(\rho),\quad T(\rho,z)=\ln{\frac{2\sqrt{\Delta_+\Delta_{-}}}{(\sqrt{\Delta_+}+\sqrt{\Delta_{-}})^2}}
\end{equation}
with $H(\rho)$ an unknown function that must be determined by means of the first equation in (\ref{mu}). Differentiating (\ref{ascia}) with respect to $\rho$ and substituting it into the first equation in (\ref{mu}) gives
\begin{equation}
\frac{dH}{d\rho}=2\rho\left[(\partial_\rho\Psi)^2-(\partial_z\Psi)^2\right]-\partial_\rho T\equiv 0,
\end{equation}
where the last step has been evaluated with Maple. Hence, $H(\rho)=c_1$ with $c_1$ an arbitrary integration constant. In order to determine $c_1$, we recall that $f\to 1$ asymptotically at space-like infinity. On the other hand, as $\rho\to\infty$  with $z$ fixed
\begin{equation}
T(\rho,z)=-\ln{2}+\mathcal{O}\left(\frac{1}{\rho^2}\right)
\end{equation}
while for $z\to\infty$ with $\rho$ fixed
\begin{equation}
T(\rho,z)=-\ln{2}+\mathcal{O}\left(\frac{1}{z^2}\right).
\end{equation}
This indicates that $c_1=\ln{2}$ and we end up with the following result
\begin{equation}
\mu+\ln{f}=\ln{\frac{4\sqrt{\Delta_+\Delta_{-}}}{(\sqrt{\Delta_+}+\sqrt{\Delta_{-}})^2}}
\end{equation}
from which it can be easily checked that $\mu\to 0$ for $\rho\to\infty$ and $z\to\infty$. Hence, our line element goes over into the Minkowski metric at space-like infinity and 
\begin{equation}
e^\mu=\frac{4\sqrt{\Delta_+\Delta_{-}}}{(\sqrt{\Delta_+}+\sqrt{\Delta_{-}})^2 f}.
\end{equation}
Finally, the metric we found is reflection symmetric with respect to the plane $z=0$ due to the $z^2$ dependence of the functions $\Delta_\pm$ and the fact that all metric coefficients are expressed in terms of such functions. Hence, our solution can be extended to the whole $z$-axis while preserving the validity of the original boundary data. Concerning the cusp singularity exhibited by the metric coefficient $g_{00}$, we compute the Kretschmann invariant $K$ by means of Maple. We find that on the equatorial plane $z=0$
\begin{equation}
K(\rho,0)=\left\{
\begin{array}{cc}
\frac{P_>(\rho)}{(\rho^2-R^2)^5} e^{-\frac{2M}{R}\arcsin{\left(\frac{R}{\rho}\right)}} & \mbox{for}~\rho>R,\\
\frac{P_<(\rho)}{4(R^2-\rho^2)^4}e^{-\pi\frac{M}{R}}&~\mbox{for}~0\leq\rho<R.
\end{array}
\right.\label{kre1}
\end{equation}
with
\begin{eqnarray}
P_>(\rho)&=&2M^2(\rho^2-R^2)^2+\left(12R^4+\frac{7}{4}M^4+2M^2 R^2\right)(\rho^2-R^2)+2M^2\left[4\rho^4+(\rho^2+R^2)^2\right]\nonumber\\
&-&M(M^2+4R^2)(\rho^2-R^2)\sqrt{\rho^2-R^2}-M\left[M^2(7\rho^2+R^2)+4R^2(R^2+3\rho^2)\right]\sqrt{\rho^2-R^2},\\
P_<(\rho)&=&7M^4+8M^2R^2+48R^4,
\end{eqnarray}
where $P_<(\rho)$ is a constant polynomial. As can be seen from Fig.~\ref{figure1as}, the Kretschmann scalar becomes infinite at $\rho=R$ on the plane $z=0$, representing a static ring-like singularity in the Weyl coordinates $(\rho,z)$. It is worth noting that the presence of such singularities in axisymmetric solutions to the Einstein field equations has been previously observed. For example, \cite{Zipoy}  demonstrated the emergence of a ring-like singularity in the equatorial plane when solving the static axisymmetric vacuum problem in oblate spheroidal coordinates. Furthermore, \cite{Semerak} discusses the notable differences between the ring-like singularities in Weyl coordinates, specifically focusing on the Bach-Weyl ring, and the ring singularity in the Kerr metric. Although the Kerr solution is not static but rather stationary, it appears to possess a simpler ring structure despite the presence of dragging effects. In contrast, the Bach-Weyl ring, which is considered analogous to the Newtonian homogeneous circular ring, exhibits directional deformations, suggesting the need for a more suitable coordinate representation and interpretation of this source. In other words, the ring singularity in the Kerr metric is relatively simpler compared to the static axisymmetric rings studied in the aforementioned paper. We would like to underline that a comprehensive analysis of the topology associated with the ring-like singularity arising from the arcsin metric is beyond the scope of our manuscript, and it would warrant a separate publication to thoroughly investigate.
\begin{figure}[!ht]\label{as}
\centering
    \includegraphics[width=0.4\textwidth]{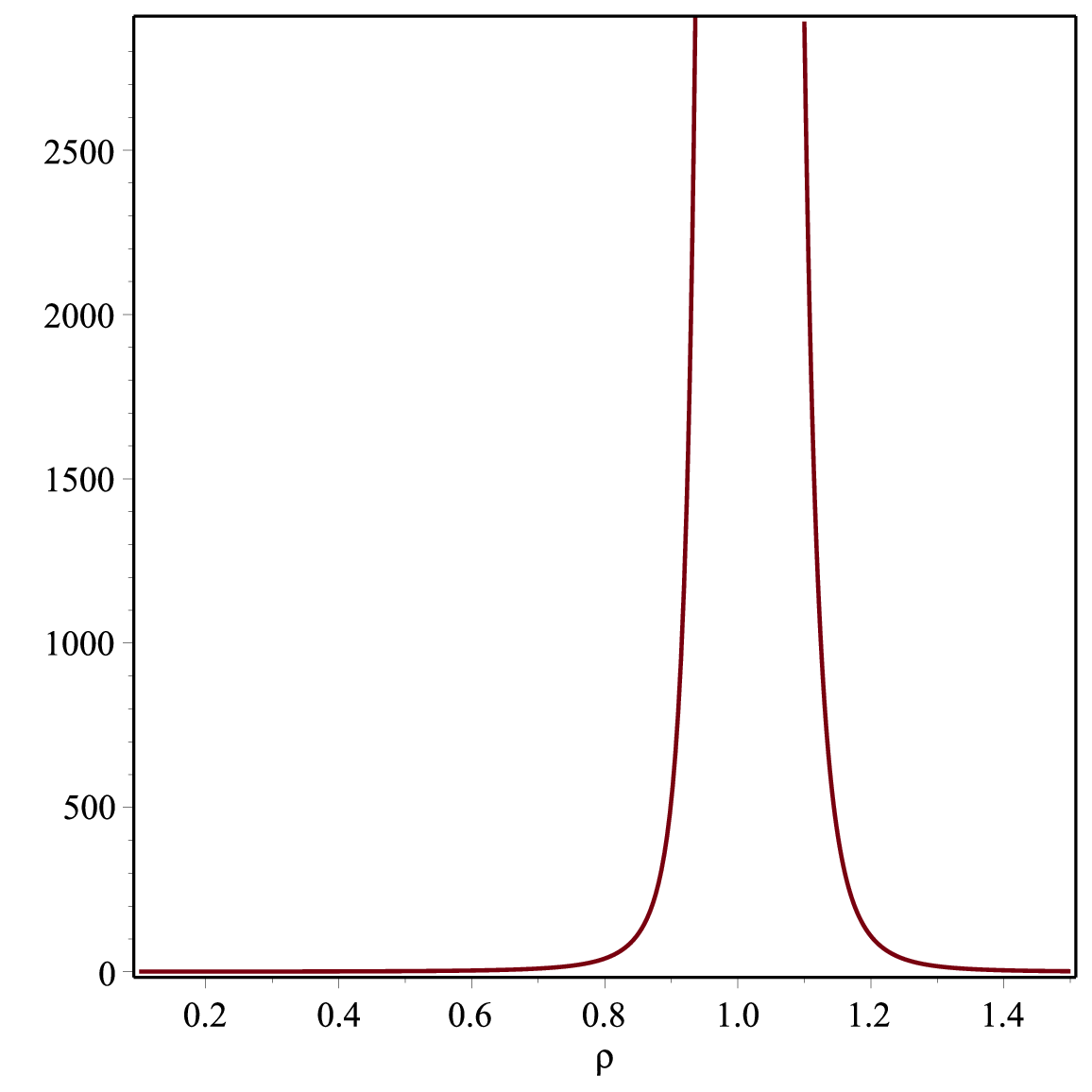}
    \includegraphics[width=0.4\textwidth]{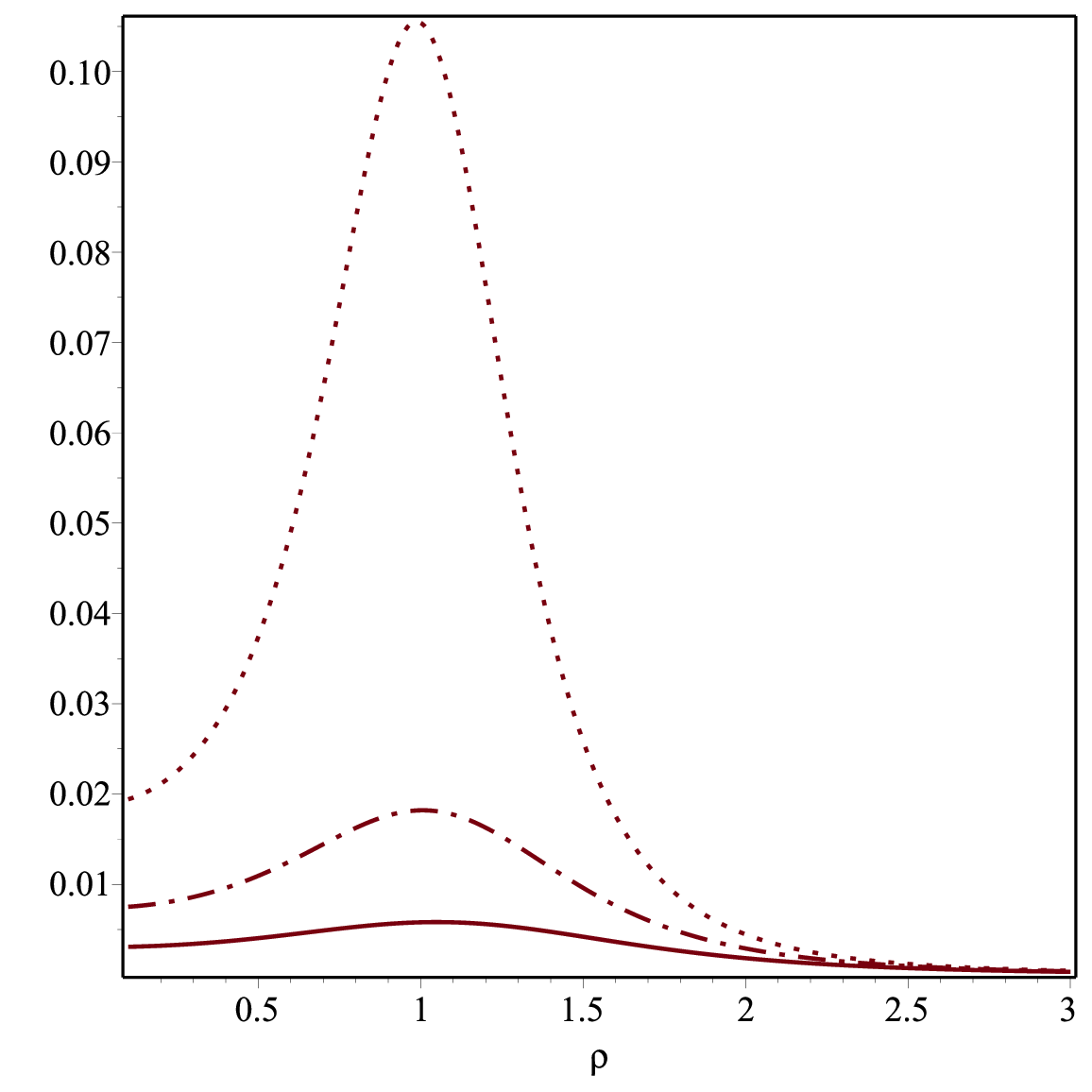}
\caption{\label{figure1as}
Plots of the Kretschmann invariant $K$ for $R=M=1$. The left panel describes the behaviour of $K$ defined in (\ref{kre1}) as a function of $\rho$ on the plane $z=0$. $K$ becomes singular at $\rho=1$. The right panel depicts $K$ as a function of $\rho$ and $z$ when $z=0.5$ (dotted line), $z=0.75$ (dash-dotted line) and $z=1$ (solid line).}
\end{figure}

\subsection{The elliptic metric}
We construct a new Weyl solution which is reflection symmetric with respect to the plane $z=0$ and reproduces the Minkowski metric at space-like infinity. To this purpose, we consider the following boundary value problem
\begin{eqnarray}
&&\frac{1}{\rho}\partial_\rho\left(\rho\partial_\rho\Psi\right)+\partial_{zz}\Psi=0\quad
\mbox{on}\quad
0<\rho<\infty,\quad 0<z<\infty,\\
&&\Psi(\rho,0)=\frac{\Psi_0}{\pi\sqrt{\gamma\rho}}Q_{-1/2}\left(\frac{\rho^2+\gamma^2}{2\gamma\rho}\right)\quad\mbox{on}\quad z=0,\\
&&g_{00}=f=e^{-2\Psi}\to 1\quad\mbox{as}\quad \rho,z\to\infty,\label{bb}
\end{eqnarray}
where $\gamma>0$ and $Q_{-1/2}$ is the Legendre function of the 2nd kind whose asymptotic behavior for large $\rho$ is \cite{Abra}
\begin{equation}
Q_{-1/2}\left(\frac{\rho^2+\gamma^2}{2\gamma\rho}\right)=\pi\sqrt{\frac{\gamma}{\rho}}+\mathcal{O}(\rho^{-5/2}).
\end{equation}
Hence, $\Psi(\rho,0)\to 0$ as $\rho\to\infty$ and the boundary condition (\ref{bb}) is trivially satisfied asymptotically on the plane $z=0$ for our initial data. Taking into account that the solution to the above boundary value problem is given by (\ref{sol_int}) and employing $6.612.3$ in \cite{Grad} immediately yield
\begin{equation}
\Psi(\rho,z)=\frac{\Psi_0}{\pi\sqrt{\gamma\rho}}Q_{-1/2}\left(\frac{\rho^2+z^2+\gamma^2}{2\gamma\rho}\right)
\end{equation}
and the $g_{00}$ metric coefficient is given by
\begin{equation}\label{gQ}
g_{00}=\mbox{exp}\left(-\frac{2\Psi_0}{\pi\sqrt{\gamma\rho}}Q_{-1/2}\left(\frac{\rho^2+z^2+\gamma^2}{2\gamma\rho}\right)\right)
\end{equation}
Note that asymptotically away $\rho^2+z^2\approx r^2$ and there, we find that 
\begin{equation}
g_{00}=1-\frac{2\Psi_0}{r}+\mathcal{O}(r^{-2}).
\end{equation}
This observation allows use to identify $\Psi_0$ as the total mass $M$ of the gravitational object associated with this space-time. As it can be seen from Figure~5, $g_{00}$ exhibits a cusp singularity along the ring $\rho=\gamma$ situated on the plane $z=0$. To understand whether this is a curvature or a coordinate singularity, it is necessary to analyse the Kretschmann scalar. To this purpose, we now derive the remaining metric coefficient $e^\mu$ entering in (\ref{Lewis-Papa}).
\begin{figure}[!ht]\label{due}
\centering
    \includegraphics[width=0.5\textwidth]{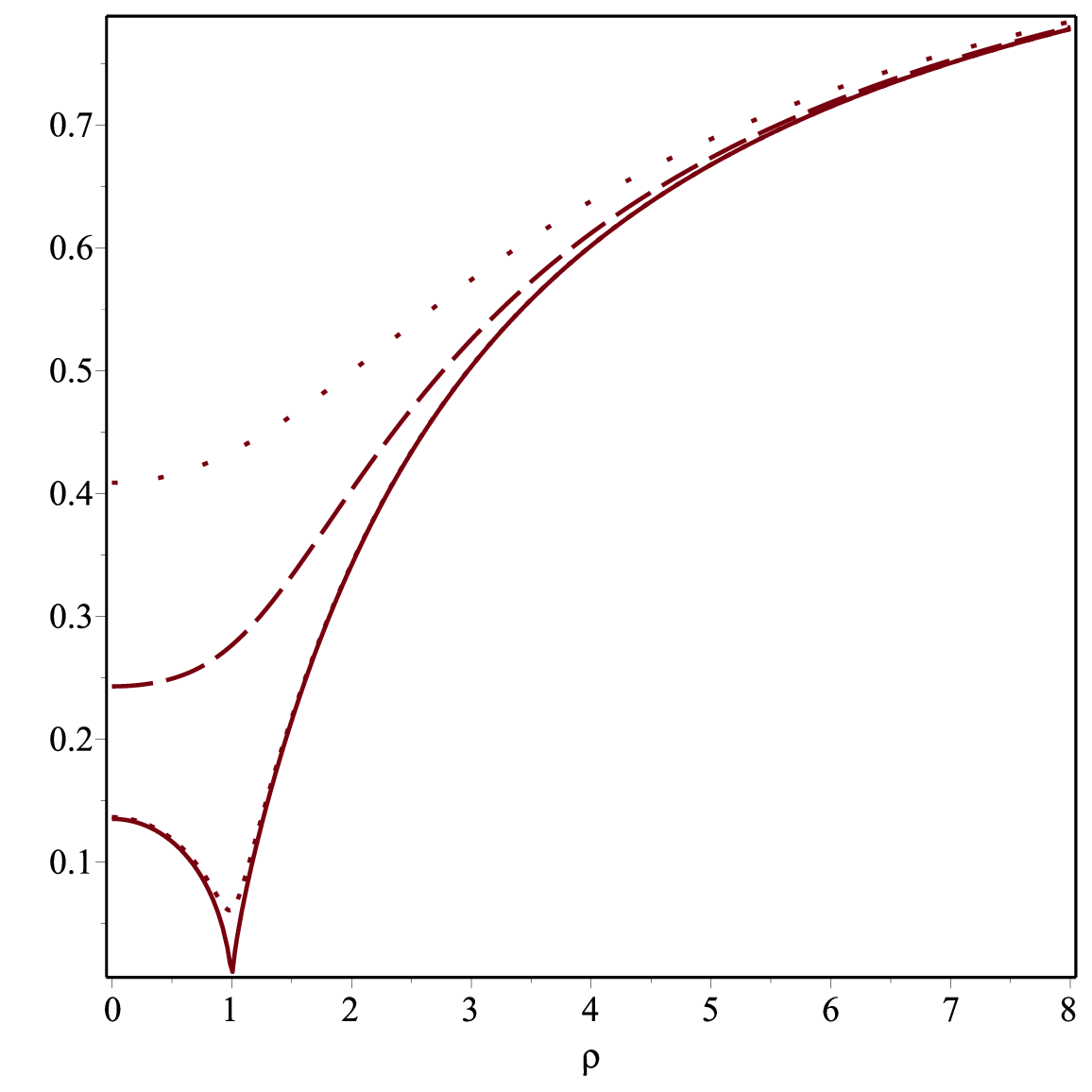}
\caption{\label{figure2}
Plot of the metric coefficient $g_{00}$ defined in (\ref{gQ}) as a function of $\rho$ for different values of $z$ and for the choice $M=1=\gamma$. The solid, dotted, dashed and space-dotted lines correspond to $z=0$, $z=0.1$, $z=1$ and $z=2$, respectively.}
\end{figure}
In order to integrate the second equation in (\ref{mu}), we need to evaluate the first order partial derivatives of $\Psi$. In this regard, it turns out to be convenient to introduce the function
\begin{equation}
h(\rho,z)=\frac{\rho^2+z^2+\gamma^2}{2\gamma\rho}.
\end{equation}
Then, the chain rule coupled to $8.732$ in \cite{Grad} gives
\begin{eqnarray}
\partial_\rho\Psi&=&-\frac{MQ_{-1/2}(h)}{2\pi\rho\sqrt{\gamma\rho}}+\frac{M\partial_\rho h}{2\pi\sqrt{\gamma\rho}(h^2-1)}\left[Q_{1/2}(h)-hQ_{-1/2}(h)\right],\label{der1}\\
\partial_z\Psi&=&\frac{M\partial_z h}{2\pi\sqrt{\gamma\rho}(h^2-1)}\left[Q_{1/2}(h)-hQ_{-1/2}(h)\right].\label{der2}
\end{eqnarray}
If we substitute (\ref{der1}) and (\ref{der2}) into the second equation in (\ref{mu}) and then, we integrate with respect to the variable $z$, we end up with 
\begin{equation}\label{intmuQ}
\mu+\ln{f}=\frac{M^2}{\pi^2\gamma}\left[-\frac{1}{\rho}\underbrace{\int F_1(\rho,z)dz}_{(I)}+\underbrace{\int F_2(\rho,z)dz}_{(II)}\right]+H(\rho),
\end{equation}
where $H$ is an unknown function and
\begin{equation}
F_1(\rho,z)=\frac{\partial_z h}{h^2-1}Q_{-1/2}(h)\left[Q_{1/2}(h)-hQ_{-1/2}(h)\right],\quad
F_2(\rho,z)=\frac{\partial_\rho h\partial_z h}{(h^2-1)^2}\left[Q_{1/2}(h)-hQ_{-1/2}(h)\right]^2.
\end{equation}
Using an identity for the Legendre functions \cite{Abra} gives for the first integral
\begin{equation}\label{primus}
(I)=2\int Q_{-1/2}(h)\frac{dQ_{-1/2}}{dh}\frac{\partial h}{\partial z}dz=Q^2_{-1/2}(h).
\end{equation}
The computation of the second integral in (\ref{intmuQ}) is more subtle. The key point here is to get rid of $\partial_\rho h$. This can be easily done by means of the identity
\begin{equation}
\partial_\rho h=\frac{1}{\gamma}-\frac{h}{\rho},
\end{equation}
which allows to break down the integral (II) as follows
\begin{eqnarray}
(II)&=&\frac{1}{\gamma}\int\frac{\partial_z h}{(h^2-1)^2}\left[Q_{1/2}(h)-hQ_{-1/2}(h)\right]^2 dz-\frac{1}{\rho}\int\frac{h\partial_z h}{(h^2-1)^2}\left[Q_{1/2}(h)-hQ_{-1/2}(h)\right]^2 dz,\\
&=&\frac{4}{\gamma}\int\left(\frac{dQ_{-1/2}}{dh}\right)^2\frac{\partial h}{\partial z}dz-\frac{1}{\rho}\int\frac{h}{(h^2-1)^2}\left[Q_{1/2}(h)-hQ_{-1/2}(h)\right]^2\frac{\partial h}{\partial z} dz,\\
&=&\frac{4}{\gamma}\underbrace{\int\left(\frac{dQ_{-1/2}}{dh}\right)^2 dh}_{(III)}-\frac{1}{\rho}\underbrace{\int\frac{h}{(h^2-1)^2}\left[Q_{1/2}(h)-hQ_{-1/2}(h)\right]^2 dh}_{(IV)},\\
\end{eqnarray}
where in the second step we used again an identity for the first derivative of a Legendre function (see \cite{Abra}). The integral (III) can be computed with Maple and we find
\begin{equation}
(III)=-\frac{1}{8(h^2-1)}\left[hQ_{1/2}(h)+hQ_{-1/2}(h)-2Q_{-1/2}(h)Q_{1/2}(h)\right].
\end{equation}
Integrating by parts (IV) gives
\begin{equation}
(IV)=-\frac{[Q_{1/2}(h)-hQ_{-1/2}(h)]^2}{2(h^2-1)}+\underbrace{\int\frac{1}{2(h^2-1)}\frac{d}{dh}[Q_{1/2}(h)-hQ_{-1/2}(h)]^2 dh}_{(V)}
\end{equation}
where (V) has been evaluated with Maple and found to be
\begin{equation}
(V)=-\frac{1}{2}Q_{-1/2}^2(h).
\end{equation}
Bringing everything together yields the following expression for the integral (II), namely
\begin{equation}\label{secundus}
(II)=-\frac{h\left[Q_{-1/2}^2(h)+Q_{1/2}^2(h)\right]-2Q_{-1/2}(h)Q_{1/2}(h)}{2\gamma(h^2-1)}+\frac{1}{2\rho}\left[Q_{-1/2}^2(h)+\frac{(Q_{1/2}(h)-hQ_{-1/2}(h))^2}{h^2-1}\right].
\end{equation}
Replacing (\ref{primus} and (\ref{secundus}) into (\ref{intmuQ}) and rearranging terms gives
\begin{equation}\label{tul}
\mu+\ln{f}=\underbrace{\frac{M^2}{2\pi^2\gamma(h^2-1)}\left[\left(\frac{1}{\rho}-\frac{h}{\gamma}\right)\left(Q_{-1/2}^2(h)+Q^2_{1/2}(h)\right)+2 Q_{-1/2}(h)Q_{1/2}(h)\partial_\rho h\right]}_{(*)}+H(\rho).
\end{equation}
As a double check of the validity of the above expression we verified numerically that the quantity $\partial_z (*)$ indeed coincides with $4\rho\partial_\rho\Psi\partial_z\Psi$. Moreover, we also checked numerically that $\partial_\rho(*)$ agrees with $2\rho\left[(\partial_\rho\Psi)^2-(\partial_z\Psi)^2\right]$ appearing in the first equation in (\ref{mu}). This signalizes that $H(\rho)\equiv 0$. Last but not least, it can be easily verified with Maple that the quantity (*) in (\ref{tul}) converges to zero as $\rho,z\to\infty$. This is gratifying because it ensures that the line element we derived does indeed reproduce the Minkowski metric asymptotically away from the gravitational source. Hence, we conclude that
\begin{equation}
e^\mu=\mbox{exp}\left(\frac{2M}{\pi\sqrt{\gamma\rho}}Q_{-1/2}(h)+\frac{M^2}{2\pi^2\gamma(h^2-1)}\left[\left(\frac{1}{\rho}-\frac{h}{\gamma}\right)\left(Q_{-1/2}^2(h)+Q^2_{1/2}(h)\right)+2 Q_{-1/2}(h)Q_{1/2}(h)\partial_\rho h\right]\right).
\end{equation}
Finally, by means of $8.13.3$ and $8.13.7$ in \cite{Abra} we can express the Legendre functions of index $\pm 1/2$ in terms of complete elliptic integrals of the first kind as follows
\begin{equation}\label{relQK}
Q_{-1/2}(h)=\widetilde{h}K(\widetilde{h}),\quad
Q_{1/2}(h)=h\widetilde{h}K(\widetilde{h})-\frac{2}{\widetilde{h}}E(\widetilde{h}),\quad
\widetilde{h}=\sqrt{\frac{4\gamma\rho}{(\rho+\gamma)^2+z^2}}.
\end{equation}
At this point, the metric coefficients can be written as
\begin{eqnarray}
g_{00}&=&\mbox{exp}\left(-\frac{4MK(\widetilde{h})}{\pi\sqrt{(\rho+\gamma)^2+z^2}}\right),\label{ell1}\\
g_{\rho\rho}&=&g_{zz}=\mbox{exp}\left(\frac{4MK(\widetilde{h})}{\pi\sqrt{(\rho+\gamma)^2+z^2}}-\frac{M^2}{\pi^2\gamma^2}\left[P_1(\rho,z)K^2(\widetilde{h})+P_2(\rho,z)E^2(\widetilde{h})-2K(\widetilde{h})E(\widetilde{h})\right]\right)\label{ell2}
\end{eqnarray}
with
\begin{equation}
P_1(\rho,z)=\frac{\rho^2+z^2+3\gamma^2}{(\rho+\gamma)^2+z^2},\quad
P_2(\rho,z)=\frac{\rho^2+z^2-\gamma^2}{(\rho-\gamma)^2+z^2}.
\end{equation}
We end our analysis with the classification of the cusp singularity of the metric coefficient $g_{00}$ on the equatorial plane. To this purpose, we used Maple to compute the Kretschmann invariant $K$ for the metric defined through (\ref{ell1}) and (\ref{ell2}). Since the corresponding analytic expression for $K$ is extremely lengthy, we decided to study $K$ numerically. As it can be seen from Table~\ref{tableEins}, 
 \begin{table}[ht]
\caption{Numerical values of the Kretschmann scalar $K$ on the equatorial plane $z=0$ for $\rho$ close to $\gamma$ when $M=\gamma=1$.}
\begin{center}
\begin{tabular}{ | l | l | l | l|}
\hline
$\rho$ & $K(\rho,0)$  \\ \hline
0.800  & 1.720432829  \\ \hline
0.850  & 1.651360590  \\ \hline
0.900  & 0.828847694  \\ \hline
0.950  & 2.864279979  \\ \hline
0.960  & 2.322546198  \\ \hline
0.970  & 0.731810061  \\ \hline
0.980  & 0.017410465  \\ \hline
0.990  & 4.3709$\cdot 10^{-9}$\\ \hline
0.995  & 1.8181$\cdot 10^{-24}$\\ \hline
1.005  & 6.1793$\cdot 10^{46}$\\ \hline
1.050  & 3.9438$\cdot 10^8$\\ \hline
1.100  & 1.5294$\cdot 10^5$\\ \hline
\end{tabular}
\label{tableEins}
\end{center}
\end{table}
$K$ blows up in proximity of $\rho=\gamma$ indicating that this is a  curvature singularity. This behaviour is also confirmed by Fig.~\ref{figurel}. In addition, we  
\begin{figure}[!ht]\label{asel}
\centering
    \includegraphics[width=0.4\textwidth]{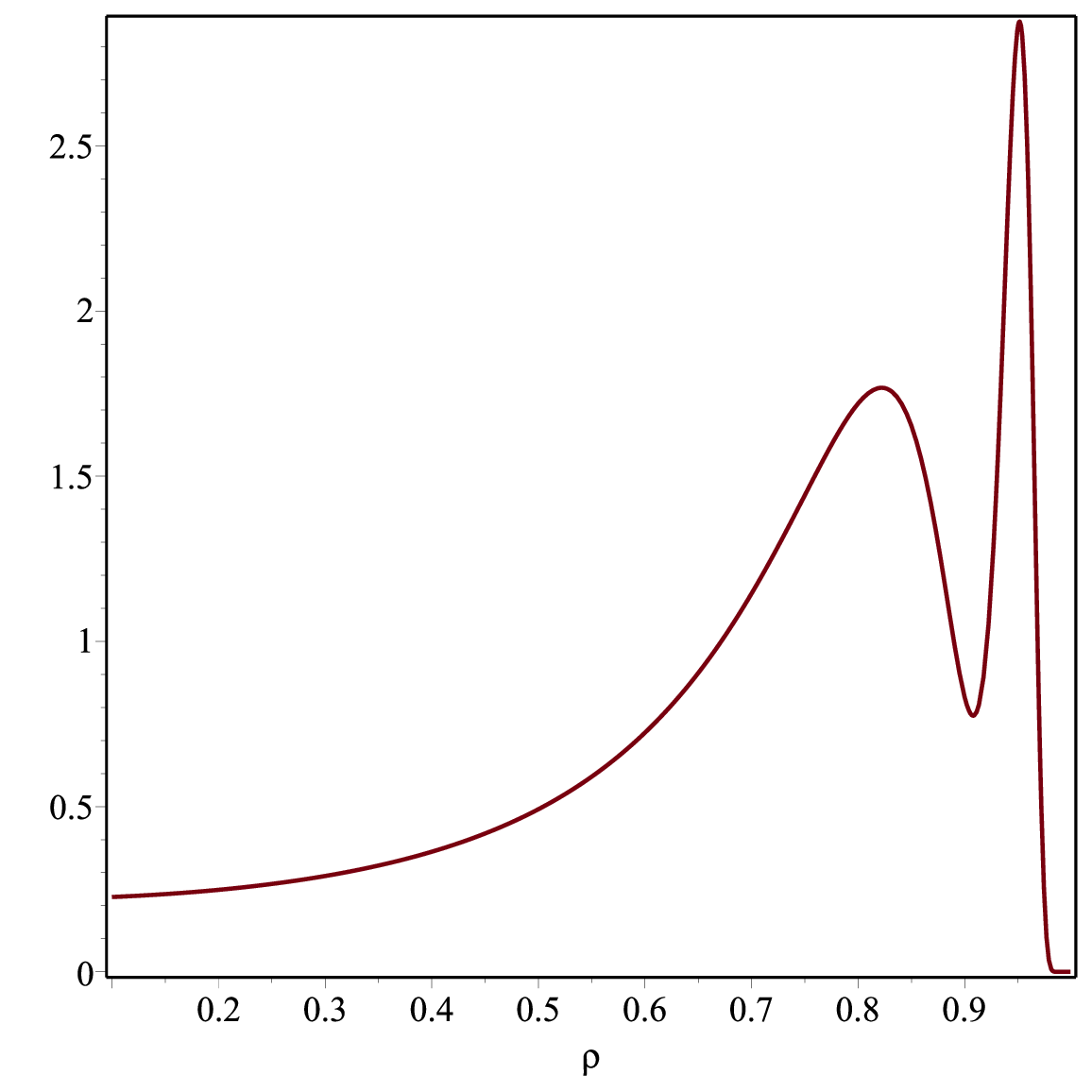}
    \includegraphics[width=0.4\textwidth]{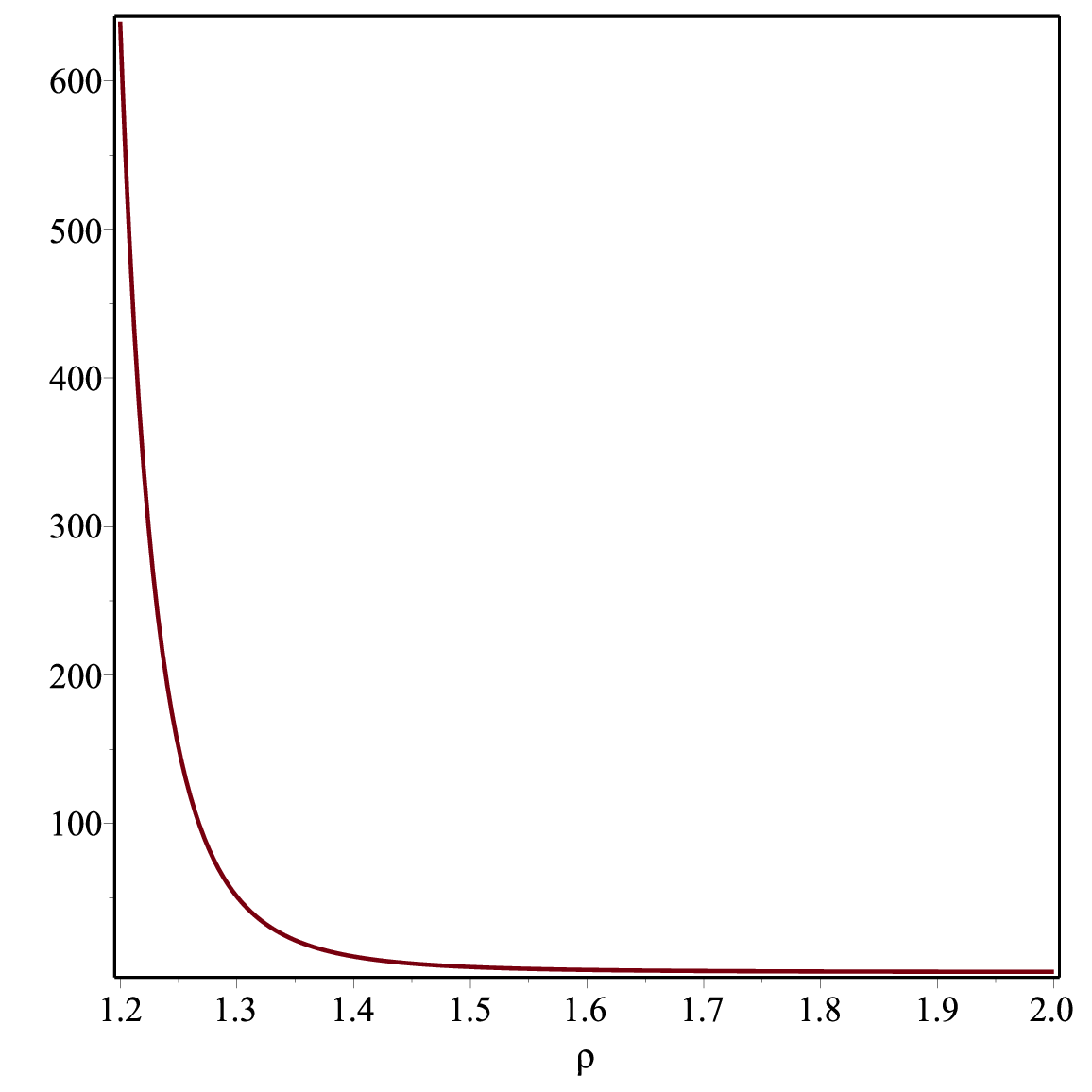}
    \includegraphics[width=0.4\textwidth]{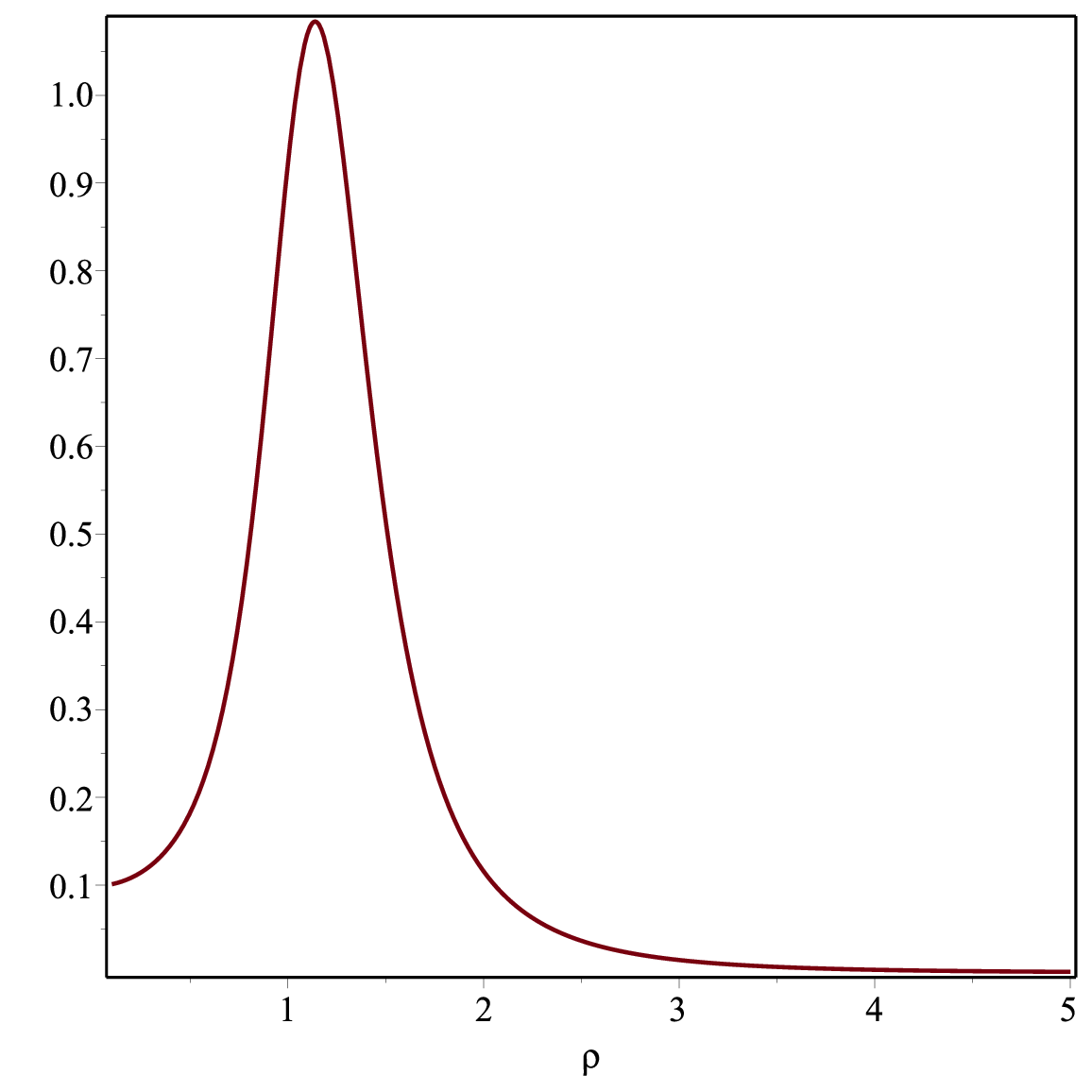}
\caption{\label{figurel}
Plots of the Kretschmann invariant $K$ for $M=\gamma=1$. The top left and right panels describe the behaviour of $K$ as a function of $\rho$ confined on the plane $z=0$. $K$ becomes singular at $\rho=1$. The bottom plot depicts $K$ on the plane $z=0.5$ where it exhibits a smooth behaviour.}
\end{figure}
observe that $K$ is regular away from the equatorial plane and takes a finite value at $\rho=0$  and $z=0$. More precisely, $K$ admits the following expansion in a neighbourhood of $\rho=0$ on the equatorial plane
\begin{equation}
K(\rho,0)=\frac{12M^2}{\gamma^6}e^{-4M/\gamma}+\mathcal{O}(\rho).
\end{equation}
If we insist in the interpretation of $\Psi$ in terms of a certain Newtonian gravitational potential, we observe that by means of (\ref{relQK}) we can bring $\Psi$ into the same form as that of a Newtonian potential for an infinitesimally thin matter coil of radius $\gamma$ (see \cite{Lass} for comparison). A quite plausible reason for the emergence of a naked singularity at $\rho=\gamma$ is that according to the above interpretation the coil cross section is zero and therefore, one would expect that the Kretchmann invariant blows up along the coil. From this perspective, the presence of the naked singularity would simply signalise the inadequacy of modelling a ring of matter in terms of a coil having zero cross section. A possible remedy might consist in  replacing the aforementioned coil with a with a finite toroidal region of matter. In this way, we would be able to account for the finite size of the ring and avoid the singularity that was present in the previous vacuum solution. The matching procedure would then allow us to combine the solutions in the two regions to obtain a complete, well-defined solution that represents the entire physical system. However, it is important to keep in mind that this technique is not without its challenges. The matching procedure can be technically challenging, especially when the two space-times have different symmetry properties or when the spacetime curvature is strong in one of the regions. Additionally, the choice of the matching surface between the two regions can have significant implications for the solution, and care must be taken to ensure that the matching is done in a physically meaningful and self-consistent way.

\section{Conclusions and outlook}
In this paper, we explore the rich landscape of axisymmetric and reflection symmetric vacuum solutions to the Einstein field equations (EFEs) using the powerful Hankel integral transform method. By applying this technique, we derive a set of new solutions that offer valuable insights into the nature of spacetime in the context of general relativity. Notably, all three solutions we obtain feature naked singularities, highlighting the presence of highly curved regions that lack the protective shield of an event horizon. These naked singularities challenge our conventional understanding of the nature of spacetime, underscoring the need for a deeper exploration of their properties and consequences. Their existence raises intriguing questions about the behavior of matter and energy in extreme gravitational environments. Furthermore, the solutions shed light on the role of axisymmetric systems and the efficacy of integral transform methods in tackling complex problems within the framework of general relativity. Through our work, we emphasize the importance of studying and understanding the behavior of singularities in the universe. The presence of naked singularities in these solutions suggests the potential for unconventional and counter-intuitive outcomes, such as extreme redshift or blueshift effects, in the surrounding spacetime. These findings motivate further research into the physical implications and astrophysical consequences of naked singularities, as well as their connection to other areas of study in general relativity and quantum gravity. We end our work by mentioning that there are several issues regarding naked singularities that are worth studying, including
\begin{itemize}
\item
{\bf{Existence}}: Determining under what conditions naked singularities can form and whether they exist in the observable world.
\item
{\bf{Stability}}: Understanding the stability of naked singularities and how they evolve over time.
\item
{\bf{Physical implications}}: Examining the physical implications of naked singularities, such as the release of large amounts of energy or radiation, and how these might affect the surrounding area.
\item
{\bf{Cosmic censorship}}: Investigating the validity of the Cosmic Censorship Hypothesis and the limitations of General Relativity.
\item
{\bf{Quantum gravity}}: Exploring the possible role of quantum gravity in resolving the issues posed by naked singularities.
\item
{\bf{Astrophysical implications}}: Studying the astrophysical implications of naked singularities, such as their potential role in the formation and evolution of galaxies and black holes.
\end{itemize}
Last but not least, our study opens up avenues for future research by highlighting the potential applications of the obtained solutions. Specifically, the arcsine and elliptic metrics exhibit characteristics that make them suitable as exterior solutions for inner regions filled with matter. Exploring the compatibility and physical implications of these solutions when coupled with appropriate matter sources is an intriguing direction for future investigations. By incorporating the dynamics of matter into the picture, we can deepen our understanding of the interplay between gravity and the distribution of energy and explore the rich possibilities that arise in such scenarios. Thus, the study of these solutions as exterior spacetimes for matter-filled regions holds great promise for uncovering new insights into the behavior of physical systems in the framework of general relativity. Future work will focus on the construction of such solutions.

\section*{Author contribution statement}
D. Batic: Conceived and designed the analysis; Analyzed and interpreted the data; Contributed analysis tools or data.\\
N. B. Debru: Analyzed and interpreted the data; Wrote the paper.\\
M. Nowakowski: Analyzed and interpreted the data; Wrote the paper.

\section*{Data availability statement}
No data was used for the research described in the article.


\begin{thebibliography}{99}
\bibitem{Birkhoff}
G. D. Birkhoff, Relativity and Modern Physics, Harvard University Press, Cambridge Massachusetts (1923).
\bibitem{Jebsen}
J. T. Jebsen, On the general spherically symmetric solutions of Einstein's gravitational equations in vacuo, Gen. Relativ. Gravit. 37 (2005) 2253.
\bibitem{Israel}
W. Isreal, Event Horizons in Static Vacuum Space-Times, Phys. Rev. 164 (1967) 5.
\bibitem{book}
H. Stephani, D. Kramer, M. MacCallum, C. Hoenselaers, H. Herlt, Exact Solutions of Einstein's Field Equations,  Cambridge University Press (2003).
\bibitem{Kerr}
R. P. Kerr, Gravitational field of a spinning mass as an example of algebraically special metrics, Phys. Rev. Lett. 11 (1963) 237. 
\bibitem{tomsato} 
A. Tomimatsu, H. Sato, New exact solution for the gravitational field of a spinning mass, Phys. Rev. Lett. 29 (1972) 1344.
\bibitem{tomsatop} 
A. Tomimatsu, H. Sato, New series of exact solutions for gravitational fields of spinning masses, Progr. Theor. Phys. 50 (1973) 95.
\bibitem{gary}
G. W. Gibbons, R. A. Russell-Clark, Note on the Tomimatsu-Sato Solution of Einstein's Equations, Phys. Rev. Lett. 30 (1973) 398.
\bibitem{DB}
D. Batic, The Tomimatsu–Sato Metric Reloaded, Universe 9 (2023) 77.
\bibitem{MP}
S. D. Majumdar, A  Class of Exact Solutions of Einstein's Field Equations, Phys. Rev. 72 (1947) 930.
\bibitem{Papas}
A Papapetrou, A Static Solution of Equations of Gravitational Field for an Arbitrary Charge Distribution, Proc. Roy. Irish Acad. A 57 (1947) 191.
\bibitem{HH}
J. B. Hartle, S. W Hawking, Solutions of Einstein-Maxwell Equations with many Black Holes, Comm. Math Phys. 26 (1971) 87.
\bibitem{Roger}
R. Penrose, Gravitational collapse: The role of general relativity, Riv. Nuovo Cim. 1 (1969) 252.
\bibitem{Sz}
P. Szekeres, Quasispherical Gravitational Collapse, Phys. Rev. D 12 (1975)) 2941.
\bibitem{dem1}
D. Christodoulou, Violation of Cosmic Censorship in the Gravitational Collapse of a Dust Cloud, Commun. Math. Phys. 93 (1984) 171.
\bibitem{dem2}
D. Christodoulou, The problem of a self-gravitating scalar field, Commun. Math. Phys. 105 (1986) 337.
\bibitem{Shapiro}
S. L. Shapiro, S. A. Teukolsky, Formation of naked singularities: The violation of cosmic censorship, Phys. Rev. Lett. 66 (1991) 994.
\bibitem{Shapiro2}
S. L. Shapiro, S. A. Teukolsky, Gravitational collapse of rotating spheroids and the formation of naked singularities, Phys. Rev. D 45 (1992) 2006.
\bibitem{Singh}
T. P. Singh, Singularities and Cosmic Censorship, J. Astrophys. Astron. 18 (1997) 335.
\bibitem{Pen}
R. Penrose, The Question of Cosmic Censorship, J. Astrophys. Astron. 20 (1999) 233.
\bibitem{Ori1}
A. Ori, T. Piran, Naked Singularities in Selfsimilar Spherical Gravitational Collapse, Phys. Rev. Lett. 59 (1987) 2137.
\bibitem{Ori2}
A. Ori, T. Piran, Naked Singularities and Other Features of Selfsimilar General Relativistic Gravitational Collapse, Phys. Rev. D 42 (1990) 1068. 
\bibitem{Joshi1}
P. S. Joshi, I. H. Dwivedi, The Structure of Naked Singularity in Self-Similar Gravitational Collapse, Commun. Math. Phys. 146 (1992) 333. 
\bibitem{Joshi2}
P. S. Joshi, I. H. Dwivedi, The Structure of naked singularity in selfsimilar gravitational collapse. 2., Lett. Math. Phys. 27 (1993) 235. 
\bibitem{Dw}
I. H. Dwivedi, P. S. Joshi, On the occurrence of naked singularity in spherically symmetric gravitational collapse, Commun. Math. Phys. 166 (1994) 117.
\bibitem{Singh1}
T. P. Singh, L. Witten, Cosmic censorship and spherical gravitational collapse with tangential pressure, Class. Quant. Grav. 14 (1997) 3489. 
\bibitem{Harada1}
T. Harada, Final fate of the spherically symmetric collapse of a perfect fluid, Phys. Rev. D 58 (1998) 104015.
\bibitem{Magli}
G. Magli, Gravitational collapse with nonvanishing tangential stresses. 2. Extension to the charged case and general solution, Class. Quant. Grav. 15 (1998) 3215. 
\bibitem{Harada2}
T. Harada, H. Iguchi, K. Nakao, Naked singularity formation in the collapse of a spherical cloud of counter rotating particles, Phys. Rev. D 58 (1998) 041502.
\bibitem{Ch1}
M. W. Choptuik, Universality and Scaling in Gravitational Collapse of a Massless Scalar Field, Phys. Rev. Lett. 70 (1992) 9.
\bibitem{Ch2}
M. W. Choptuik, Critical behaviour in a scalar field collapse. In: D. Hobill, A. Burd, A. Coley (eds) {\it{Deterministic Chaos in General Relativity}}. NATO ASI Series 332, Springer Verlag, Boston, MA (1994).
\bibitem{Hamade}
R. S. Hamade, J. M. Stewart, The spherically symmetric collapse of a massless scalar field, Class. Quant. Grav. 13 (1996) 497.
\bibitem{Joshi3}
P. S. Joshi, N. Dadhich, R. Maartens, Why do naked singularities form in gravitational collapse?, Phys. Rev. D 65 (2002) 101501.
\bibitem{Joshi4}
P. S. Joshi, A. Krolak, Naked strong curvature singularities in Szekeres space-times, Class. Quant. Grav. 13 (1996) 3069. 
\bibitem{Sandip}
S. K. Chakrabarti, P. S. Joshi, Naked Singularities as Possible Candidates for Gamma-ray Bursters, Int. J. Mod. Phys. D 3 (1994) 647.
\bibitem{Maeda}
H. Maeda, Final fate of spherically symmetric gravitational collapse of a dust cloud in Einstein-Gauss-Bonnet gravity, Phys. Rev. D 73 (2006) 104004.
\bibitem{Amir}
A. H. Ziaie, K. Atazadeh , Y. Tavakoli, Naked Singularity Formation In Brans-Dicke Theory, Class. Quant. Grav. 27 (2010) 075016; erratum ibidem: Class. Quant. Grav. 27 (2010) 209801.
\bibitem{Rudra}
P. Rudra, R. Biswas, U. Debnath, Gravitational Collapse In Husain Space-time For Brans-Dicke Gravity Theory with Power-law Potential, Astrophys. Space Sci. 354 (2014) 597.
\bibitem{Gibbons1}
G. W. Gibbons, S. A. Hartnoll, A. Ishibashi, On the stability of naked singularities, Prog. Theor. Phys. 113 (2005) 963.
\bibitem{Gleiser}
R. J. Gleiser, G. Dotti, Instability of the negative mass Schwarzschild naked singularity, Class. Quant. Grav. 23 (2006) 5063.
\bibitem{Cardoso}
V. Cardoso, M. Cavaglia, Stability of naked singularities and algebraically special modes, Phys. Rev. D 74 (2006) 024027.
\bibitem{Dotti1}
G. Dotti, R. J. Gleiser, J. Pullin, Instability of charged and rotating naked singularities, Phys. Lett. B 644 (2007) 289.
\bibitem{Dotti2}
G. Dotti, R. J. Gleiser, The Initial value problem for linearized gravitational perturbations of the Schwarzschild naked singularity, Class.  Quant. Grav. 26 (2009) 215002.
\bibitem{Dotti3}
G. Dotti, R. J. Gleiser, Gravitational instability of the inner static region of a Reissner-Nordstr\"{o}m black hole, Class. Quant. Grav. 27 (2010) 185007.
\bibitem{Dotti4}
G. Dotti, Linear Stability of Black Holes and Naked Singularities,  Universe 8 (2022) 38.
\bibitem{Lewis}
T. Lewis, Some Special Solutions of the Equations of Axially Symmetric Gravitational Fields, Proc. R. Soc. London A 136 (1932) 176.
\bibitem{Papa}
A. Papapetrou, Eine rotationssymmetrische L\"{o}sung in der allgemeinen Relativit\"{a}tstheorie, Annals Phys. 12 (1953) 309.
\bibitem{Ernst}
F. J. Ernst, New Formulation of the Axially Symmetric Gravitational Field Problem, Phys. Rev. 167 (1968) 1175.
\bibitem{parrymoon}
P. Moon, D. Eberle Spencer, Field theory handbook, Springer Berlin, Heidelberg (1988).
\bibitem{Carmeli}
M. Carmeli, Classical fields: General relativity and gauge theory, World Scientific Publishing (2001).
\bibitem{GP}
J. B. Griffiths, J. Podolsky, Exact Space-Times in Einstein's General Relativity, Cambridge University Press (2012).
\bibitem{Futa}
T. Futamase, B. Schutz, Newtonian and post-Newtonian approximations are asymptotic to general relativity, Phys. Rev. D 28 (1983) 2363.
\bibitem{coop1}
F. I. Cooperstock, S. Tieu, General relativity resolves galactic rotation without exotic dark matter,  astro-ph/0507619 (2005).
\bibitem{coop2}
F. I. Cooperstock and S. Tieu, Galactic dynamics via general relativity: a compilation and new developments, Int. J. Mod. Phys. 22 (2007) 2293.
\bibitem{Carr}
J. Carrick and F. Cooperstock, General relativistic dynamics applied to the rotation curves of galaxies, Astrophys. Space Sci. 337 (2012) 321.
\bibitem{Caccia}
D. Astesiano, S. L. Cacciatori, V. Gorinin, F. Re, Towards a full general relativistic approach to galaxies, Eur. Phys. J. C 82 (2022) 554.
\bibitem{gw}
B. P. Abbott et al., Observation of gravitational waves from a binary black hole merger, Phys. Rev. Lett. 116 (2016) 061102.
\bibitem{Carlotto}
A. Carlotto, R. Schoen, Localizing solutions of the Einstein constraint equations, Invent. Math. 205 (2015) 559.
\bibitem{Anderson}
P. Anderson, D. Brill, Gravitational geons revisited, Phys. Rev. D 56 (1997) 4824.
\bibitem{Deb}
L. Debnath, D. Bhatta, Integral Transforms and their applications, 2nd edition, Chapmann \& Hall (2007).
\bibitem{SHK}
R. Shaikh, S. Kar, Wormholes, the weak energy condition, and scalar-tensor gravity, Phys. Rev. D 94 (2016) 024011.
\bibitem{Curzon}
H. Curzon, Cylindrical solutions of Einstein’s gravitation equations, Proc. Lond. Math. Soc. 23 (1925) 477.
\bibitem{Grad}
I. S. Gradshteyn, I. M. Ryzhik, Table of Integrals, Series, and Products, Academic Press (2007).
\bibitem{Gaut}
R. Gautreau, J. L. Anderson, Directional singularities in Weyl gravitational fields, Phys. Lett. A 25 (1967) 291.
\bibitem{Stac}
J. Stachel, Structure of the Curzon metric, Phys. Lett. A 27 (1968) 60.
\bibitem{CJ}
F. I. Cooperstock, G. J. Junevicus, Singularities in Weyl gravitational fields, Int. J. Theor. Phys. 9 (1974) 59.
\bibitem{SM}
P. Szekeres, F. H. Morgan, Extensions of the Curzon metric, Comm. Math. Phys. 32 (1973) 313.
\bibitem{SS}
S. M. Scott, P. Szekeres, The abstract boundary - a new approach to singularities of manifolds, J. Geom. Phys. 13 (1994) 223.
\bibitem{Lind}
L. Lindblom, On the symmetries of equilibrium stellar models, Phil. Trans. R. Soc. Lond. A 340 (1992) 353.
\bibitem{Meinel}
R. Meinel, G. Neugebauer, Asymptotically flat solutions to the Ernst equation with reflection symmetry, Class. Quant. Grav. 12 (1995) 2045.
\bibitem{Luke}
Y. L. Luke, Integrals of Bessel Functions, McGraw-Hill Book Company (1962).
\bibitem{Abra}
M. Abramowitz, I. Stegun, Handbook of Mathematical Functions with Formulas, Graphs, and Mathematical Tables, Dover, New York (1964).
\bibitem{Zipoy}
D. M. Zipoy, Topology of some spheroidal metrics, J. Math. Phys. 7 (1966) 1137.
\bibitem{Semerak}
O. Semer$\acute{\mbox{a}}$k, Static axisymmetric rings in general relativity: How diverse they are, Phys. Rev. D 94 (2016) 104021.
\bibitem{Lass}
H. Lass, L. Blitzer, The gravitational potential due to uniform disks and rings, Celest. Mech. 30 (1983) 225.
\end{thebibliography}
\end{document}